\documentclass[leqno,draft,11pt,a4paper]{article}
\usepackage{amssymb,amsmath,latexsym,theorem}
\usepackage[includeheadfoot,margin=1in]{geometry}

\usepackage[final]{graphicx}

\allowdisplaybreaks[2]

\newcommand\nequiv{\not\equiv}
\newcommand\LOR{\bigvee}
\newcommand\ET{\bigwedge}
\newcommand\nto{\nrightarrow}
\newcommand\dual{\mathrm d}

\newcommand\fii{\varphi}
\newcommand\ep{\varepsilon}

\newcommand\p[1]{\langle#1\rangle}
\newcommand\lh[1]{\lvert#1\rvert}
\newcommand\Abs[1]{\left|#1\right|}
\newcommand\bez{\smallsetminus}
\newcommand\sset{\subseteq}
\newcommand\nsset{\nsubseteq}
\newcommand\Sset{\supseteq}
\newcommand\pw[1]{\mathcal P(#1)}
\newcommand\nul{\varnothing}
\newcommand\two{\mathbf2}

\newcommand\fl[1]{\lfloor#1\rfloor}
\newcommand\cl[1]{\lceil#1\rceil}
\newcommand\fdiv{\genfrac\lfloor\rfloor{}{}}
\newcommand\cdiv{\genfrac\lceil\rceil{}{}}

\DeclareMathOperator\PR{Pr}
\DeclareMathOperator\inv{Inv}
\DeclareMathOperator\pol{Pol}
\DeclareMathOperator\gr{gr}

\newcommand\cxt[1]{\mathbf{#1}}
\newcommand\np{\cxt{NP}}
\newcommand\conp{\cxt{coNP}}
\newcommand\ptime{\cxt P}
\newcommand\EXP{\cxt{EXP}}
\newcommand\nci{\cxt{NC}^1}
\newcommand\psp{\cxt{PSPACE}}
\newcommand\tc{\cxt{TC}^0}
\newcommand\Ac{\cxt{AC}^0}
\newcommand\tpt{\cxt{\Theta}^\ptime_2}
\newcommand\task[1]{{\normalfont\textsc{#1}}}
\newcommand\cmp{\task{CMP}}
\newcommand\Eq{\task{EQ}}

\newcommand\N{\mathbb N}
\newcommand\GF[1]{\mathbb F_{#1}}
\newcommand\pres{\triangleright}
\newcommand\npres{\ntriangleright}
\newcommand\Op{\mathrm{Op}}
\newcommand\Rel{\mathrm{Rel}}
\newcommand\cln[1]{\mathrm{#1}}

\newcommand\txto{${}\to{}$}
\newcommand\paren[1]{\textup(\nobreak\hskip0pt\relax#1\textup)}

\DeclareRobustCommand*\legend[1]{\lower1pt \hbox{\includegraphics[width=8bp]{clonememb-leg-#1}}}

\newcommand\noproof{\leavevmode\unskip\bme\vadjust{}\nobreak\hfill$\Box$\par}
\newcommand\bme{\hskip.75em\relax}
\newenvironment{Pf}
  {\par\noindent\textit{Proof:}\bme\ignorespaces}
  {\noproof\pagebreak[2]\vskip\medskipamount\ignorespacesafterend}

\theoremstyle{plain}
\newtheorem{Thm}{Theorem}[section]
\newtheorem{Cor}[Thm]{Corollary}
\newtheorem{Lem}[Thm]{Lemma}
\newtheorem{Fact}[Thm]{Fact}
\newtheorem{Prob}[Thm]{Problem}
\theorembodyfont\upshape
\newtheorem{Rem}[Thm]{Remark}

\author{Emil Je\v r\'abek\\[\medskipamount]
The Czech Academy of Sciences, Institute of Mathematics\\
\small \v Zitn\'a 25,
115\:67 Praha 1,
Czech Republic,
email: \texttt{jerabek@math.cas.cz}
}

\title{On the complexity of the clone membership problem}

\begin{document}
\maketitle

\begin{abstract}
We investigate the complexity of the Boolean clone membership problem ($\cmp$): given a set of Boolean functions~$F$ and
a Boolean function~$f$, determine if $f$ is in the clone generated by~$F$, i.e., if it can be expressed by a circuit
with $F$-gates. Here, $f$ and elements of~$F$ are given as circuits or formulas over the usual De~Morgan basis.
B\"ohler and Schnoor~\cite{boeh-schn} proved that for any fixed~$F$, the problem is $\conp$-complete, with a few
exceptions where it is in~$\ptime$. Vollmer~\cite{voll:basis} incorrectly claimed that the full problem $\cmp$ is also
$\conp$-complete. We prove that $\cmp$ is in fact $\tpt$-complete, and we complement B\"ohler and Schnoor's
results by showing that for fixed~$f$, the problem is $\np$-complete unless $f$ is a projection.

More generally, we study the problem $B$-$\cmp$ where $F$ and~$f$ are given by circuits using gates from~$B$. For most
choices of~$B$, we classify the complexity of $B$-$\cmp$ as being $\tpt$-complete (possibly under randomized
reductions), $\cxt{coDP}$-complete, or in~$\ptime$.
\end{abstract}

\section{Introduction}\label{sec:introduction}

The fundamental concept of a \emph{clone} originated in universal algebra as an abstraction of the notion of an
algebra---a clone consists of operations term-definable in a given algebra; it reappears in logic in the form of a
collection of truth functions (in a possibly multi-valued logic) definable by formulas over a given set of connectives,
and in computer science as a set of functions computable by circuits over a given basis of gates. Clones have many
applications in theoretical computer science in classification of the complexity of CSPs~\cite{schaef,bul:csp,zhuk:csp}
or satisfiability and related problems~\cite{lew:sat,reith:phd}, including various nonclassical logics
\cite{bmsssv:ltl,bmtv:deflt,cst:abduc,mei-sch:alc}.
The most basic computational problem associated with clones is the \emph{clone membership problem}, asking if a given
function is expressible (definable by a term, formula, or circuit) from a given set of initial functions.
This problem has computer science applications in the context of formal algebraic specifications~\cite{san-tar:obs-eq}.

From the point of view of computational complexity, several variants of the clone membership problem were studied in
the literature. The most straightforward representation of the input functions is by tables of values. In this setting,
Kozen~\cite{kozen:pspace} proved that the membership problem for clones of \emph{unary} functions on arbitrary finite
domains is $\psp$-complete. The general clone membership problem for arbitrary functions on finite domains is
$\EXP$-complete. This result is credited in~\cite{berg-slut} to an unpublished manuscript of H.~Friedman, but the first
published proof is due to Bergman, Juedes, and Slutzki~\cite{berg-jued-slut}; a mistake in their paper was corrected by
Ma\v sulovi\'c~\cite{masul:corr}. As shown by Kozik~\cite{kozik:exp}, there even exists a \emph{fixed} finitely
generated clone on a finite domain whose membership problem is $\EXP$-complete.

The high complexity of the problem on arbitrary finite domains is related to the complicated structure of clones on
domains of size $\ge3$. In contrast, the lattice of clones on the \emph{Boolean} (two-element) domain is quite simple,
and it has been explicitly described by Post~\cite{post}. As a result, the Boolean clone membership problem is
computationally much easier than the general case: it was shown to be in~$\cxt{NL}$ by Bergman and
Slutzki~\cite{berg-slut}, while Vollmer~\cite{voll:basis} proved that it was in quasipolynomial $\Ac$, which implies it
is not $\cxt{NL}$-hard (or even $\Ac[2]$-hard).

The representation of functions by tables is quite inefficient, as it always has size exponential in the number of
variables. A viable alternative, especially in the Boolean case, is to represent functions by expressions (circuits or
formulas) over some canonically chosen functionally complete basis, say, the De~Morgan basis $\{\land,\lor,\neg\}$.
In this setting, B\"ohler and Schnoor~\cite{boeh-schn} studied the complexity of membership problems for
\emph{fixed} Boolean clones~$C$: they proved that all such problems are $\conp$-complete with a few exceptions that are
in~$\ptime$. More generally, they studied variants of the problem where $f$ is not given by a circuit over a
functionally complete basis, but over an arbitrary (but fixed) basis. They classified most such problems as being
$\conp$-complete or in~$\ptime$.

The full Boolean clone membership problem in the circuit representation (denoted $\cmp$ in this paper) was considered
by Vollmer~\cite{voll:basis}, who claimed it was also $\conp$-complete. However, he did not provide much in the way of
proof for the $\conp$ upper bound (see Remark~\ref{rem:voll} below), and as we will see, this claim is wrong.

A characterization of clone membership in terms of preservation of relational invariants easily implies that $\cmp$ is
computable in $\ptime^\np$---more precisely, in the class $\tpt=\ptime^{\np[\log]}=\ptime^{\|\np}$. The main goal of this paper is to prove that $\cmp$ is in fact $\tpt$-complete. As a
warm-up, we consider a restriction of $\cmp$ dual to B\"ohler and Schnoor's results: we prove that for a fixed target
Boolean function~$f$, the clone membership problem is $\np$-complete in all nontrivial cases (i.e., unless $f$ is a
projection function, or a nullary function if we allow them). This already shows that $\cmp$ cannot be $\conp$-complete
unless $\np=\conp$. We then go on to prove that $\cmp$ is $\tpt$-complete;
our main technical tool is a characterization of clones generated by threshold functions.
We also discuss some variants of our results, such as using formulas instead of circuits for representation of
functions, or allowing nullary functions.

In the second part of the paper, we investigate the complexity of restricted versions of $\cmp$, denoted $B$-$\cmp$,
where the input functions are given by circuits or formulas over an arbitrary (but fixed) finite basis~$B$ instead of
the De~Morgan basis. We show that $B$-$\cmp$ remains $\tpt$-complete, albeit using randomized reductions, when the
clone $[B]$ generated by~$B$ has infinitely many subclones, and includes some non-monotone functions; we rely on a
randomized construction of formulas for threshold functions using fixed threshold functions as gates, following the
method of Valiant~\cite{val:maj}. On the other hand, if $[B]$ has only finitely many subclones, we classify the
complexity of $B$-$\cmp$ as either $\cxt{coDP}$-complete or in~$\ptime$. The complexity of $B$-$\cmp$ remains open when
$[B]$ has infinitely many subclones, but consists of monotone functions only.

\section{Preliminaries}\label{sec:preliminaries}

\subsection{Boolean functions and clones}\label{sec:bool-funct-clon}

We will assume basic familiarity with the theory of Boolean clones; this is described in many places, for example
Lau~\cite{lau}. We will summarize the most important points below to fix our terminology and notation.

Let $\two=\{0,1\}$. An \emph{$n$-ary Boolean function} (or \emph{operation}) is a mapping $f\colon\two^n\to\two$. We
denote the set of $n$-ary Boolean functions by~$\Op_n$, and the set of all Boolean functions by
$\Op=\bigcup_{n\ge1}\Op_n$. (Following the tradition in literature on Boolean clones, we disallow nullary functions; we
will comment later on how this affects our results.)

We will use common connectives such as $\land,\lor,\to,\neg$ to denote specific Boolean functions; by abuse of notation,
$0$ and~$1$ denote the constant functions of arbitrary arity. If $0\le i<n$, the \emph{projection function}
$\pi^n_i\in\Op_n$ is defined by $\pi^n_i(x_0,\dots,x_{n-1})=x_i$. For any $n>0$ and $0\le t\le
n+1$, the \emph{threshold function} $\theta^n_t\in\Op_n$ is defined by
\[\theta^n_t(x_0,\dots,x_{n-1})=1\iff\bigl|\{i<n:x_i=1\}\bigr|\ge t.\]
Notice that $\theta^n_0=1$, $\theta^n_{n+1}=0$, $\theta^2_1=\lor$, $\theta^2_2=\land$, and $\theta^1_1=\pi^1_1$ is the
identity function.

Given functions $f\in\Op_n$ and $g_0,\dots,g_{n-1}\in\Op_m$, their \emph{composition} is the function $h\in\Op_m$
defined by
\[h(\vec x)=f\bigl(g_0(\vec x),\dots,g_{n-1}(\vec x)\bigr).\]
A set of Boolean functions $C\sset\Op$ is a \emph{clone} if it contains all projections and is closed under
composition. The intersection of an arbitrary collection of clones is again a clone (where the empty intersection is
understood as~$\Op$), thus the poset of clones under inclusion forms a complete lattice,
and it yields an (algebraic) closure operator $[-]\colon\pw\Op\to\pw\Op$; that is, for any $F\sset\Op$, $[F]$ denotes
the clone generated by~$F$.

The \emph{dual} of a Boolean function $f$ is the function
$f^\dual(x_0,\dots,x_{n-1})=\neg f(\neg x_0,\dots,\neg x_{n-1})$. The dual of a clone~$C$ is the clone
$\{f^\dual:f\in C\}$.

A \emph{$k$-ary Boolean relation} is $r\sset\two^k$. The set of $k$-ary Boolean relations is denoted by $\Rel_k$,
and the set of all relations by $\Rel=\bigcup_k\Rel_k$. (Here, it will not make any difference if we allow nullary
relations or not; the reader is welcome to make the choice.) A function $f\in\Op_n$ \emph{preserves} a relation
$r\in\Rel_k$, written as $f\pres r$, if $f$, considered as a mapping of the relational structures
$\p{\two,r}\times\dots\times\p{\two,r}\to\p{\two,r}$, is a homomorphism. Explicitly, $f\pres r$ iff the following
implication holds for every matrix $(a_i^j)_{i<n}^{j<k}\in\two^{k\times n}$:
\[\begin{pmatrix}a_0^0\\\vdots\\a_0^{k-1}\end{pmatrix}\in r,\dots,
  \begin{pmatrix}a_{n-1}^0\\\vdots\\a_{n-1}^{k-1}\end{pmatrix}\in r
  \implies
  \begin{pmatrix}f(a_0^0,\dots,a_{n-1}^0)\\\vdots\\f(a_0^{k-1},\dots,a_{n-1}^{k-1})\end{pmatrix}\in r.\]

For example, the relation $r=\{\p{x,y}:x\lor y=1\}$ is not preserved by the function $f(x,y,z)=x\land(y\lor z)$, as
witnessed by the matrix $\bigl(\begin{smallmatrix}1&0&0\\0&1&1\end{smallmatrix}\bigr)$: every column is in~$r$, but
applying $f$ row-wise gives the vector $\bigl(\begin{smallmatrix}0\\0\end{smallmatrix}\bigr)\notin r$. On the other
hand, $\theta^3_2$ preserves $r$ (and all other $2$-ary relations for that matter): for any $(a_i^j)_{i<3}^{j<2}$ and
each $j<2$, $\theta^3_2(a_0^j,a_1^j,a_2^j)\ne a_i^j$ for at most one $i<3$, hence there exists $i<3$ such that
\[\Bigl(\begin{smallmatrix}\theta^3_2(a_0^0,a_1^0,a_2^0)\\\theta^3_2(a_0^1,a_1^1,a_2^1)\end{smallmatrix}\Bigr)=
\Bigl(\begin{smallmatrix}a_i^0\\a_i^1\end{smallmatrix}\Bigr).\]

If $F\sset\Op$ and $R\sset\Rel$, we write $F\pres R$ if $f\pres r$ for all $f\in F$ and $r\in R$. The set of
\emph{invariants} of~$F\sset\Op$ and the set of \emph{polymorphisms} of~$R\sset\Rel$ are defined by
\begin{align*}
\inv(F)&=\{r\in\Rel:F\pres r\},\\
\pol(R)&=\{f\in\Op:f\pres R\}.
\end{align*}
We will use the following fundamental properties of the $\inv$ and~$\pol$ operators (see \cite[Thms.\ 2.6.1,
2.9.1]{lau}):
\begin{Fact}\label{fact:galois}
The mappings $\inv\colon\pw\Op\to\pw\Rel$ and $\pol\colon\pw\Rel\to\pw\Op$ form an antitone Galois connection:
$R\sset\inv(F)\iff F\sset\pol(R)$.

The Galois-closed subsets of~$\Op$ are exactly the clones: i.e., $\pol(R)$ is a clone for every $R\sset\Rel$, every
clone is of the form $\pol(R)$ for some~$R\sset\Rel$, and $[F]=\pol(\inv(F))$ for all $F\sset\Op$. 
\end{Fact}

If we allowed nullary functions, then Galois-closed subsets of~$\Rel$ would be exactly the \emph{coclones}:
subsets $R\sset\Rel$ that are closed under definitions by \emph{primitive positive} formulas, i.e., first-order
formulas $\fii(\vec x)$ of the form $\exists\vec y\,\ET_{i<k}\psi_i(\vec x,\vec y)$, where each $\psi_i$ is an atomic
formula (an instance of a relation $r\in R$, or of equality). Under our restriction to non-nullary functions,
Galois-closed subsets of~$\Rel$ are only the coclones that contain the empty relation $\nul\in\Rel_1$. (See
Behrisch~\cite{behr:null} for a detailed discussion.)

The lattice of Boolean clones was completely described by Post~\cite{post}. In particular, we will make use of the
following characterization (which follows from \cite[pp.~36ff]{szen:clo} or \cite[Thm.~3.1.1]{lau}), fixing our naming
of basic clones and their invariants along the way. Here, for any $f\in\Op_n$,
$\gr(f)=\{\p{\vec x,y}\in\two^{n+1}:y=f(\vec x)\}\in\Rel_{n+1}$ denotes the graph of~$f$. Recall that an element $a$ of
a lattice~$L$ is \emph{meet-irreducible} if $a=\bigcap S$ implies $a\in S$ for any finite $S\sset L$, where $\bigcap$
denotes the lattice meet operation; if this holds also for infinite $S\sset L$, it is \emph{completely
meet-irreducible}.
\begin{Fact}\label{fact:meet-irr}
Every Boolean clone is an intersection of a family of completely meet-irreducible clones, which are:
\begin{itemize}
\item The clone $\cln M=[\land,\lor,0,1]=\pol({\le})$ of \emph{monotone} functions, where
$\le$ denotes the relation $\{\p{0,0},\p{0,1},\p{1,1}\}\in\Rel_2$.
\item The clone $\cln A=[+,1]=\pol(r_A)$ of \emph{affine} functions, where $+$ denotes addition in the two-element
field~$\GF2$ (i.e., XOR), and $r_A=\{\p{x,y,z,w}:x+y+z+w=0\}\in\Rel_4$.
\item The clone $\cln D=[\theta^3_2,\neg]=\pol\bigl(\gr(\neg)\bigr)$ of \emph{self-dual} functions \paren{i.e.,
functions $f$ such that $f=f^\dual$}.
\item The clone $\cln\ET=[\land,0,1]=\pol\bigl(\gr(\land)\bigr)$ of \emph{conjunctive} functions.
\item The clone $\cln\LOR=[\lor,0,1]=\pol\bigl(\gr(\lor)\bigr)$ of \emph{disjunctive} functions.
\item The clone $\cln U=[\neg,0]=\pol(r_U)$ of \emph{essentially unary} functions \paren{i.e., functions that depend on
at most one variable}, where $r_U=\bigl\{\p{x,y,z}:z\in\{x,y\}\bigr\}\in\Rel_3$.
\item For each $m\ge1$, the clones $\cln T^m_1=[\theta^{m+1}_2,\to]=\pol(r^m_1)$ and
$\cln T^m_0=[\theta^{m+1}_m,\nto]=\pol(r^m_0)$, where $x\nto y=\neg(x\to y)=x\land\neg y$, and the relations
$r^m_\alpha\in\Rel_m$ are defined by
\begin{align*}
r^m_1&=\{\vec x\in\two^m:x_0\lor\dots\lor x_{m-1}=1\},\\
r^m_0&=\{\vec x\in\two^m:x_0\land\dots\land x_{m-1}=0\}.
\end{align*}
Since $r^1_\alpha=\{\alpha\}$, this includes as a special case the clones $\cln P_1=[\land,\to]=\cln T^1_1$ and
$\cln P_0=[\lor,\nto]=\cln T^1_0$ of \emph{$1$-preserving} and \emph{$0$-preserving} functions \paren{respectively}.
\end{itemize}
\end{Fact}
We will denote intersections of named clones by juxtaposition, so that, e.g., $\cln{AD}=\cln A\cap\cln D$. For
convenience, we also put $\cln P=\cln P_0\cln P_1$ and $\cln T^\infty_\alpha=\bigcap_m\cln T^m_\alpha$. We have $\cln
T^\infty_1=[\to]$ and $\cln T^\infty_0=[\nto]$. We will denote the top and bottom of the lattice of clones by
$\cln\top$ and~$\cln\bot$, i.e., $\cln\top=\Op$, and $\cln\bot=\{\pi^n_i:i<n\}=\cln{UP}$. We define
\[R_n=\{{\le},r_A,\gr(\neg),\gr(\land),\gr(\lor),r_U\}\cup\{r^m_\alpha:0<m\le n,\alpha\in\two\}\sset\Rel\]
for each $n\ge0$, and $R_\infty=\bigcup_nR_n$.

\begin{figure}[t]
\centering
\includegraphics[width=300bp]{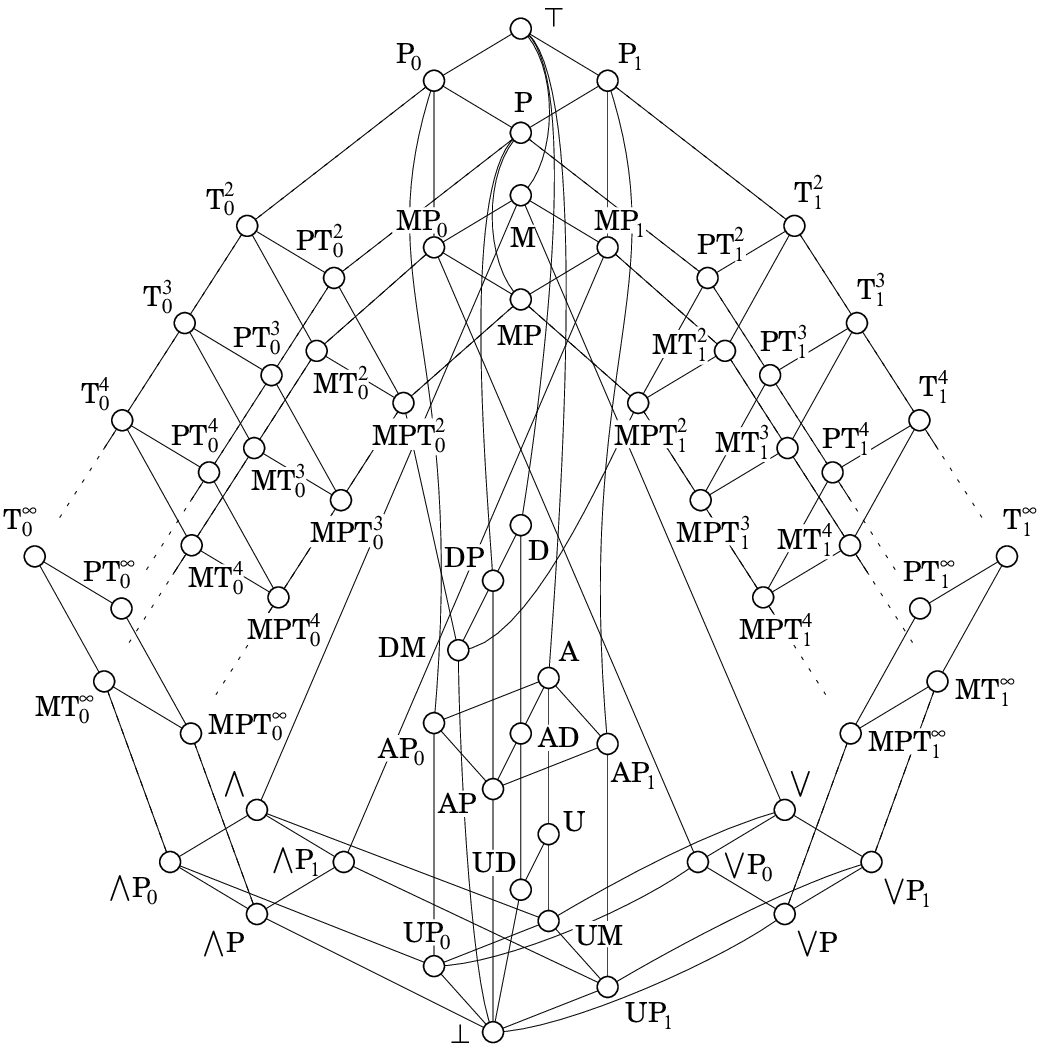}
\caption{The lattice of Boolean clones (Post's lattice).}
\label{fig:post}
\end{figure}
The Hasse diagram of the lattice of Boolean clones (called \emph{Post's lattice}) is depicted in Fig.~\ref{fig:post}.
(In fact, Post~\cite{post} did not work with the modern definition of clones, but with a slightly weaker concept of
\emph{iterative classes}, which do not necessarily contain all projections. Thus, his original lattice has four more
classes.)

\subsection{Computational complexity}\label{sec:comp-compl}

We assume the reader is familiar with basic notions of complexity theory, including the classes $\ptime$, $\np$,
and~$\conp$, and logspace functions.

% If $C$ is a complexity class, we say that a language~$L$ is \emph{$C$-hard} if for every $L'\in C$, there
% exists a many-one (uniform) $\tc$~reduction from $L'$ to~$L$; if, moreover, $L\in C$, then $L$ is \emph{$C$-complete}.
% We chose $\tc$~reductions because they strike the right balance for our purposes. On the one hand, they are fairly
% restrictive: not only are they stricter than log-space or poly-time reductions that are often used to define
% $\np$-completeness, but they also have less power than our other classes of interest, $\nci$ and~$\ptime$, hence they
% give rise to a sensible notion of $\nci$-completeness. On the other hand, $\tc$ is powerful enough to support the kind
% of syntactic manipulations that we will use to define our reductions, such as substituting one formula into another. We
% believe that the bulk of our completeness results actually hold under more restricted notions of reductions, such as
% dlogtime reductions, but the extra effort needed to get there would distract us from the main point of this paper.

If $C$ is a complexity class, we say that a language~$L$ is \emph{$C$-hard} if for every $L'\in C$, there exists a
logspace many-one reduction from $L'$ to~$L$; if, moreover, $L\in C$, then $L$ is \emph{$C$-complete}. A reader
familiar with uniform $\tc$~functions (see e.g.\ \cite{founif}) may easily find that all our results below actually
hold with $\tc$ many-one reductions in place of logspace reductions. (We believe that the bulk of our results actually
hold under even more restricted notions of reductions such as dlogtime or $\Ac$, but this might require extra effort,
or possibly a relaxed syntactic representation of circuits and formulas allowing for insertion of padding.)

Let $B\sset\Op$ be finite. A \emph{(Boolean) circuit over a basis~$B$} (or \emph{$B$-circuit} for short) in
$n$~variables is a labelled directed acyclic graph~$C$ (possibly with multiple edges) with exactly one node of
out-degree~$0$ (the output node), such that each node of~$C$ is either labelled with one of the variables $x_i$, $i<n$,
in which case it has in-degree~$0$, or it is labelled with a $k$-ary function $f\in B$, in which case it has
in-degree~$k$, and it comes with an explicit numbering of the incoming edges (such a node is called an
\emph{$f$-gate}). A circuit computes a function $\two^n\to\two$ in the usual way. A function $f\in\Op$ is
computable by a $B$-circuit iff $f\in[B]$. Given a $B$-circuit $C\colon\two^n\to\two$ and an assignment $a\in\two^n$,
we can evaluate $C(a)$ in polynomial time; moreover, evaluation of circuits over the \emph{De~Morgan basis}
$\{\land,\lor,\neg\}$ is $\ptime$-complete (Ladner~\cite{lad:circval}). The \emph{depth} of a circuit is the maximal
length of a path from a leaf to the output node.

A \emph{$B$-formula} is a $B$-circuit such that all nodes have out-degree at most~$1$. Formulas are usually represented
as strings in a more economical way than directed graphs: the \emph{prefix} (\emph{Polish}) notation can be defined
inductively such that each variable $x_i$ is a $B$-formula, and if $f\in B$ is $k$-ary, and $\fii_0,\dots,\fii_{k-1}$
are $B$-formulas, then $f\fii_0\dots\fii_{k-1}$ is a $B$-formula. If all $f\in B$ have arity at most~$2$, we may also
use the common \emph{infix} notation where $(\fii_0\mathbin f\fii_1)$ is employed instead of $f\fii_0\fii_1$ for $f$ of
arity~$2$. Evaluation of formulas over any finite basis is in~$\nci$, and over the De~Morgan basis it is
$\nci$-complete under dlogtime reductions (Buss~\cite{buss:bfvp}). Here, $\nci$ is the class of languages computable in
time $O(\log n)$ on an alternating Turing machine (with random-access input); equivalently, it consists of languages
computable by a dlogtime-uniform (more precisely, $U_E$-uniform) sequence of De~Morgan circuits of depth $O(\log
n)$, see Ruzzo~\cite{ruzzo} for details. We have $\nci\sset\cxt L$.

By duplicating nodes as needed, we can unravel any $B$-circuit of depth~$d$ into a $B$-formula of depth~$d$, and
therefore of size $2^{O(d)}$; in particular, any circuit of depth $O(\log n)$ can be transformed into a polynomial-size
formula. One consequence of this is that we can efficiently convert formulas of depth $O(\log n)$ to a different basis
(the corresponding result for circuits being trivial), which will be useful in Section~\ref{sec:restr-input-bases}:
\begin{Lem}\label{lem:conv}
Let $B,B'\sset\Op$ be finite sets such that $B\sset[B']$.
\begin{enumerate}
\item\label{item:7}
Given a $B$-circuit, we can construct in logspace an equivalent $B'$-circuit.
\item\label{item:8}
Given a $B$-formula of depth $O(\log n)$, we can construct in logspace an equivalent $B'$-formula.
\end{enumerate}
\end{Lem}
\begin{Pf}

\ref{item:7}: For each $f\in B$, we fix an expression of $f$ in terms of~$B'$, and replace with it all $f$-gates in the
circuit.

\ref{item:8}: Starting from a $B$-formula of depth $O(\log n)$, the construction above produces a $B'$-circuit of depth
$O(\log n)$, which can be unravelled into a $B'$-formula of depth $O(\log n)$, and consequently of size $n^{O(1)}$.
\end{Pf}

The \emph{Boolean clone membership problem $\cmp$} is the following decision problem:
\begin{quote}
{\bf Input:} A finite set of functions $F\sset\Op$ and a function $f\in\Op$, all given by Boolean circuits over the
De~Morgan basis.

{\bf Output:} YES if $f\in[F]$, otherwise NO.
\end{quote}
For any clone $C=[C]$, the membership problem $\cmp_C$ is the special case of~$\cmp$ where $F$ is fixed:
\begin{quote}
{\bf Input:} A function $f\in\Op$, given by a Boolean circuit over the De~Morgan basis.

{\bf Output:} YES if $f\in C$, otherwise NO.
\end{quote}
Dually, for a fixed function $f\in\Op$, $\cmp^f$ denotes the following special case of~$\cmp$:
\begin{quote}
{\bf Input:} A finite set of functions $F\sset\Op$, given by Boolean circuits over the De~Morgan basis.

{\bf Output:} YES if $f\in[F]$, otherwise NO.
\end{quote}
Notice that $\cmp^f=\cmp^g$ whenever $[f]=[g]$, as the output condition can be stated as $[f]\sset[F]$.

For any class~$C$ and $q\colon\N\to\N$, $\ptime^{C[q]}$ denotes the class of languages computable in
polynomial time using at most $q(n)$ queries to an oracle $A\in C$; similarly for a set of such functions~$q$. The
class $\ptime^{\|C}$ consists of languages computable in polynomial time with non-adaptive (parallel) access to an
oracle $A\in C$; we also write $\ptime^{\|C[q]}$ if the number of oracle queries is bounded by~$q(n)$.

The class $\tpt$, introduced by Wagner~\cite{wagner:theta2}, is defined as $\ptime^{\np[O(\log n)]}$. It has several other
equivalent characterizations, see Buss and Hay~\cite{buss-hay}: in particular, $\tpt=\ptime^{\|\np}$.

The \emph{Boolean hierarchy $\cxt{BH}$} is defined as the smallest class that includes~$\np$, and is closed under
(finitary) intersections, unions, and complements; alternatively, it can be characterized as
$\cxt{BH}=\ptime^{\np[O(1)]}$. The Boolean hierarchy is stratified into levels: the bottom level consists of $\np$ and
$\conp$, and the next level of the classes
\begin{align*}
\cxt{DP}&=\{L_0\cap L_1:L_0\in\np,L_1\in\conp\},\\
\cxt{coDP}&=\{L_0\cup L_1:L_0\in\np,L_1\in\conp\}.
\end{align*}
We will use the observation that for any $\np$-complete language~$L_0$ and a $\conp$-complete language~$L_1$, the
language
\[L=\{\p{x,y}:x\in L_0\text{ or }y\in L_1\}\]
is $\cxt{coDP}$-complete: given a $\cxt{coDP}$~language $L'=L'_0\cup L'_1$, where $L'_0\in\np$ and
$L'_1\in\conp$, we may fix reductions $f_i$ of $L'_i$ to~$L_i$ for $i=0,1$; then $w\mapsto\p{f_0(w),f_1(w)}$ is a
reduction of $L'$ to~$L$.

\section{The complexity of $\cmp$}\label{sec:complexity-cmp}

In this section, we will investigate the complexity of $\cmp$ and its subproblems $\cmp_C$ and $\cmp^f$. We will start
with upper bounds ($\cmp\in\tpt$, $\cmp_C\in\conp$, $\cmp^f\in\np$) in the first subsection. The corresponding lower
bounds will follow in Section~\ref{sec:lower-bounds-1inp} for $\cmp_C$ and~$\cmp^f$, and in
Section~\ref{sec:lower-bound-cmp} for~$\cmp$ (the $\tpt$-completeness of~$\cmp$ will be the main result of this
section). We discuss how these results are affected if we vary the definition of $\cmp$ and its subproblems in
Section~\ref{sec:alternative-setups}.

\subsection{Upper bounds}\label{sec:upper-bounds}

Similarly to \cite{berg-slut,boeh-schn,voll:basis}, we will extract an algorithm for~$\cmp$ from Fact~\ref{fact:galois},
which implies
\begin{equation}\label{eq:15}
f\in[F]\iff\inv(F)\sset\inv(f)\iff\forall r\in\Rel\:(F\npres r\text{ or }f\pres r).
\end{equation}
Using Fact~\ref{fact:meet-irr}, we may restrict attention to $r\in R_\infty$, but this is still an infinite number of
invariants due to the two infinite families $r_0^m$ and~$r_1^m$, $m\in\N$, which correspond to the two infinite arms of
Post's lattice. In order to turn~\eqref{eq:15} into an algorithmic criterion, we need to place an efficient bound
on~$m$ in terms of $F$ and~$f$, which is the content of the next lemma.
\begin{Lem}\label{lem:tma}
Let $n\ge1$, $f\in\Op_n$, and $\alpha\in\two$. The following are equivalent:
\begin{enumerate}
\item\label{item:1} $f\in\cln T^\infty_\alpha$.
\item\label{item:2} $f\in\cln T^n_\alpha$.
\item\label{item:3} There is $i<n$ such that $x_i\le^\alpha f(\vec x)$ for all $\vec x\in\two^n$, where
${\le}^1={\le}$, ${\le}^0={\ge}$.
\end{enumerate}
\end{Lem}
\begin{Pf}
\ref{item:1}\txto\ref{item:2} is trivial.

\ref{item:3}\txto\ref{item:1}: Assume $\alpha=1$ and $x_{i_0}\le f(\vec x)$. If $m\in\N$ and $(a_i^j)_{i<n}^{j<m}$ is
such that $\LOR_{j<m}a_i^j=1$ for all $i<n$, then in particular $\LOR_{j<m}a_{i_0}^j=1$, hence
$\LOR_{j<m}f(a^j_0,\dots,a^j_{n-1})=1$ by assumption.

\ref{item:2}\txto\ref{item:3}: If \ref{item:3} does not hold, let us fix for each $i<n$ a vector
$a^i=\p{a^i_0,\dots,a^i_{n-1}}\in\two^n$ such that $a^i_i=\alpha$ and $f(a^i)=\neg\alpha$. Then the matrix
$(a_i^j)_{i<n}^{j<n}$ witnesses that $f\npres r^n_\alpha$: if, say, $\alpha=1$, we have
$a_i^0\lor\dots\lor a_i^{n-1}\ge a_i^i=1$ for each $i<n$, but $f(a^0)\lor\dots\lor f(a^{n-1})=0$.
\end{Pf}
\begin{Cor}\label{cor:Rn}
If $F\sset\Op$ and $f\in\Op_n$, then
\begin{equation}\label{eq:3}
f\in[F]\iff\forall r\in R_n\:(F\npres r\text{ or }f\pres r).
\end{equation}
\end{Cor}
\begin{Pf}
The left-to-right implication is clear. For the right-to-left implication, if $f\notin[F]$, then one of the completely
meet-irreducible clones $C$ as given in Fact~\ref{fact:meet-irr} satisfies $F\sset C$ and $f\notin C$. Moreover, if
$C=\cln T^m_\alpha$ for some $\alpha\in\two$ and $m>n$, then also $f\notin\cln T^n_\alpha\Sset C\Sset F$ by
Lemma~\ref{lem:tma}. Thus, we may assume $C=\pol(r)$ for some $r\in R_n$, i.e., $F\pres r$ and $f\npres r$.
\end{Pf}

Since the criterion in~\eqref{eq:3} uses the preservation relation $\pres$, we need to know its complexity.
\begin{Lem}\label{lem:pres-conp}
If $f\in\Op$ and $r\in\Rel$ are given by Boolean circuits over any fixed finite basis, we can test whether $f\pres r$
in~$\conp$.
\end{Lem}
\begin{Pf}
Given circuits computing $f\colon\two^n\to\two$ and $r\sset\two^k$ (where we may assume w.l.o.g.\ that all variables
appear in the circuits), we have straight from the definition that $f\npres r$ iff there exists a matrix
$A=(a_i^j)_{i<n}^{j<k}\in\two^{k\times n}$ such that $\p{a_i^0,\dots,a_i^{k-1}}\in r$ for each $i<n$, and
\[\p{f(a^0_0,\dots,a^0_{n-1}),\dots,f(a^{k-1}_0,\dots,a^{k-1}_{n-1})}\notin r.\]
Here, $A$ is an object of size polynomial in $n$ and~$k$, which are bounded by the size of the input, and the stated
properties of~$A$ can be checked in polynomial time.
\end{Pf}

We can combine Corollary~\ref{cor:Rn} and Lemma~\ref{lem:pres-conp} to obtain the desired upper bounds.
\begin{Thm}\label{thm:ub}
\ \begin{enumerate}
\item\label{item:4} $\cmp\in\tpt$.
\item\label{item:5} $\cmp_C\in\conp$ for each clone $C\sset\Op$.
\item\label{item:6} $\cmp^f\in\np$ for each $f\in\Op$.
\end{enumerate}
\end{Thm}
\begin{Pf}

\ref{item:4}: Given $F\sset\Op$ and $f\in\Op$ of arity~$n$, we can determine if $f\in[F]$ in $\ptime^{\|\np}$ by
evaluating~\eqref{eq:3}: there are $2n+O(1)$ relations in~$R_n$, and they can be described by
efficiently constructible Boolean circuits. Thus, in view of Lemma~\ref{lem:pres-conp}, we can ask the $\np$~oracle if
$F\pres r$ and if $f\pres r$ for each $r\in R_n$ in parallel, and read the answer off of the oracle responses.

\ref{item:5}: We use Corollary~\ref{cor:Rn} again, but since $C$ is fixed, we can test $C\pres r$ in deterministic
polynomial time: $\{r\in R_0:C\pres r\}$ is a finite set, and for each $\alpha\in\two$, $\{m\in\N_{>0}:C\pres
r^m_\alpha\}$ is a downward-closed subset of~$\N_{>0}$, i.e., either all of~$\N_{>0}$, or a finite set. Thus,
\eqref{eq:3} can be evaluated in~$\conp$.

\ref{item:6} is even simpler: since $f$ (hence $n$) is fixed, $R_n$ is a fixed finite set, and so is $\{r\in
R_n:f\npres r\}$. Thus, we can test if $F\npres r$ for each $r$ from this finite set in~$\np$ by Lemma~\ref{lem:pres-conp}.
\end{Pf}
\begin{Rem}\label{rem:voll}
The purported proof of $\cmp\in\conp$ in Vollmer~\cite{voll:basis} essentially just states that it
follows from a criterion similar to our Corollary~\ref{cor:Rn}, despite that it involves both positive and negative
occurrences of the $\conp$~preservation relation.
\end{Rem}

\subsection{Hardness of $\cmp_C$ and $\cmp^f$}\label{sec:lower-bounds-1inp}

We turn to lower bounds which will show that Theorem~\ref{thm:ub} is mostly optimal, with a few exceptions in
\ref{item:5} and~\ref{item:6}. We start with $\cmp_C$ and $\cmp^f$, where we can prove $\cxt{(co)NP}$-hardness by
simple reductions from Boolean satisfiability.

The case of $\cmp_C$ was settled by B\"ohler and Schnoor~\cite{boeh-schn}. For completeness, we include a
self-contained proof, employing a reduction function that we will reuse in the argument for~$\cmp^f$ below.
\begin{Thm}[B\"ohler and Schnoor]\label{thm:cmp-C}
Let $C$ be a Boolean clone.
\begin{enumerate}
\item\label{item:36}
If $C\Sset\cln P$, then $\cmp_C\in\ptime$. More precisely, $\cmp_\cln\top$
is trivial, and $\cmp_C$ is $\ptime$-complete for $C=\cln P_0,\cln P_1,\cln P$.
\item\label{item:37}
Otherwise \paren{i.e., if $C\sset\cln M$, $\cln D$, $\cln A$, $\cln T^2_0$, or $\cln T^2_1$}, $\cmp_C$ is $\conp$-complete.
\end{enumerate}
\end{Thm}
\begin{Pf}

\ref{item:36}: We have $f\in\cln P_\alpha$ iff $f(\alpha,\dots,\alpha)=\alpha$, which can be checked in polynomial
time. To show the $\ptime$-hardness of $\cmp_\cln P$ and $\cmp_{\cln P_1}$, let a De~Morgan circuit~$C(x_0,\dots,x_{n-1})$ and an
assignment~$\vec a$ be given. Put $C_{\vec a}(x)=x\land C(x^{a_0},\dots,x^{a_{n-1}})$, where $x^1=x$, $x^0=\neg x$.
Then $C(\vec a)=1$ iff $C_{\vec a}\in\cln P_1$ iff $C_{\vec a}\in\cln P$. The case of $\cln P_0$ is dual.

\ref{item:37}: That $C$ is included in $\cln M$, $\cln D$, $\cln A$, $\cln T^2_0$, or $\cln T^2_1$ follows by
inspection of Post's lattice (Fig.~\ref{fig:post}). We have $\cmp_C\in\conp$ by Theorem~\ref{thm:ub}. In order to show
$\conp$-hardness, we will provide a reduction from $\task{UnSat}$; it will even be independent of~$C$.

Given a formula $\fii$ in variables $\vec u$, let
\[f_\fii(x,y,z,\vec u)=\bigl((x\land\fii)\land y\bigr)\lor\bigl(\neg(x\land\fii)\land z\bigr).\]
Then
\begin{alignat}3
\label{eq:1}\fii&\in\task{UnSat}&&\implies f_\fii\equiv z&&\implies f_\fii\in\cln\bot\sset C,\\
\label{eq:2}\fii&\in\task{Sat}&&\implies f_\fii\notin\cln M,\cln A,\cln D,\cln T^2_0,\cln T^2_1&&\implies f_\fii\notin C.
\end{alignat}
Indeed, \eqref{eq:1} is obvious. For~\eqref{eq:2}, if $\vec a$ is a satisfying assignment to~$\fii$, we see that
\[(x\land y)\lor(\neg x\land z)=f_\fii(x,y,z,\vec a)\in[f_\fii,0,1],\]
thus $f_\fii\notin\cln M,\cln A$. Moreover,
\begin{align*}
f_\fii(1,1,0,\vec a)&=f_\fii(0,0,1,\neg\vec a)=1,\\
f_\fii(1,0,1,\vec a)&=f_\fii(0,1,0,\neg\vec a)=0,
\end{align*}
thus $f_\fii\notin\cln D,\cln T^2_1,\cln T^2_0$.
\end{Pf}

Using a similar construction, we can now determine the complexity of~$\cmp^f$.
\pagebreak[2]
\begin{Thm}\label{thm:cmp-f}
Let $f\in\Op$.
\begin{enumerate}
\item If $f$ is a projection \paren{i.e., $f\in\cln\bot$}, then $\cmp^f$ is trivial.
\item\label{item:39}
Otherwise, $\cmp^f$ is $\np$-complete.
\end{enumerate}
\end{Thm}
\begin{Pf}
Let $\fii\mapsto f_\fii$ be the reduction from the proof of Theorem~\ref{thm:cmp-C}. It follows from the definition that
$f_\fii\in\cln P$, while \eqref{eq:2} implies that $[f_\fii]\Sset\cln P$ if $\fii$ is satisfiable. Thus,
\begin{alignat*}2
\fii&\in\task{Sat}&&\implies[f_\fii]=\cln P,\\
\fii&\in\task{UnSat}&&\implies[f_\fii]=\cln\bot,
\end{alignat*}
hence $\fii\mapsto\{f_\fii\}$ is a reduction from $\task{Sat}$ to $\cmp^f$ whenever $f\in\cln P\bez\cln\bot$. Likewise,
the reduction $\fii\mapsto\{f_\fii,0,1\}$ works whenever $f\notin[0,1]$, and $\fii\mapsto\{f_\fii,\neg\}$ works
whenever $f\notin[\neg]$. This covers all cases where $f$ is not a projection.
\end{Pf}

Since $\cmp^f$ is a special case of~$\cmp$, Theorem~\ref{thm:cmp-f} also shows the $\np$-hardness of~$\cmp$, which
(conditionally) refutes Vollmer's claim $\cmp\in\conp$.
\begin{Cor}\label{cor:not-conp}
$\cmp\notin\conp$ unless $\np=\conp$.
\noproof\end{Cor}

It will take more work to establish the true complexity of $\cmp$, which we will do in the next subsection.

\subsection{Hardness of $\cmp$}\label{sec:lower-bound-cmp}

We aim to prove that $\cmp$ is $\tpt$-complete in this subsection. First, if we look at how the $\tpt$ upper bound was
derived in the proof of Theorem~\ref{thm:ub}, we see that the only way the complexity of~$\cmp$ can get as hard as $\tpt$ is
by interaction of $F$ and~$f$ deep inside one of the infinite arms of Post's lattice: otherwise \eqref{eq:3} holds with
a constant~$n$, in which case the criterion can be evaluated in $\ptime^{\np[O(1)]}$, i.e., in the Boolean hierarchy
(cf.\ Lemma~\ref{lem:b-cmp-bh}). Thus, we need a flexible supply of functions that generate clones on the infinite arms:
we will find a suitable pool of such functions among the threshold functions~$\theta^n_t$. We then consider a
suitable $\tpt$-hard problem based on counting the number of satisfiable formulas from a given list, which we transform
into a question about the clone generated by a certain formula that combines the list using a threshold function.

In view of the outline above, our first task is to describe exactly the clones generated by threshold functions. For
completeness, the lemma below is stated including various cases that we will not actually need.
\begin{Lem}\label{lem:thr}
Let $n\ge1$ and $0\le t\le n+1$. Then
\[[\theta^n_t]=\begin{cases}
\cln\bot&t=n=1,\\
\cln{UP}_1&t=0,\\
\cln{{\LOR}P}&t=1,\quad n>1,\\
\cln{MPT}_1^{\fl{(n-1)/(t-1)}}&1<t\le\frac n2,\\
\cln{DM}&t=\frac{n+1}2,\quad n>1,\\
\cln{MPT}_0^{\cl{t/(n-t)}}&\frac n2+1\le t<n,\\
\cln{{\ET}P}&t=n,\quad n>1,\\
\cln{UP}_0&t=n+1.
\end{cases}\]
\end{Lem}
\begin{Pf}
The cases with $t\le1$ or $t\ge n$ are straightforward.

Notice that the dual of~$\theta^n_t$ is $\theta^n_{n+1-t}$. Thus, if $t=(n+1)/2$ (which implies $n$ is odd), then
$\theta^n_t\in\cln{DM}$. By Fig.~\ref{fig:post}, $\cln{DM}$ is a minimal clone, hence $[\theta^n_t]=\cln{DM}$ unless
$\theta^n_t$ is a projection, which only happens when $n=1$.

Assume that $1<t\le n/2$. Clearly, $\theta^n_t\in\cln{MP}$. Since $\N_{<n}=\{0,\dots,n-1\}$ has two disjoint subsets of
size $\fl{n/2}\ge t$, $\theta^n_t\notin\pol(r^2_0)=\cln T^2_0$. Also, $t\ge2$ implies that $\theta^n_t$ is not bounded
below by a variable, i.e., $\theta^n_t\notin\cln T^\infty_1$ by Lemma~\ref{lem:tma}. By inspection of Post's lattice, it
follows that $[\theta^n_t]=\cln{MPT}^k_1$ for some $k\ge1$. Now, for any $k\ge1$, we have $\theta^n_t\npres r^k_1$ iff
$\N_{<n}$ can be covered by $k$ subsets of size $<t$ iff $n\le k(t-1)$, thus
\[\theta^n_t\in\cln T^k_1\iff n\ge1+k(t-1)\iff k\le\fdiv{n-1}{t-1},\]
and consequently $[\theta^n_t]=\cln{MPT}_1^{\fl{(n-1)/(t-1)}}$.

Finally, assume that $\frac n2+1\le t<n$. The dual of $\theta^n_t$ is $\theta^n_{t'}$, where $t'=n+1-t$ satisfies
$1<t'\le\frac n2$. Thus, by the case that we just proved, $[\theta^n_{t'}]=\cln{MPT}^k_1$ with
\[k=\fdiv{n-1}{t'-1}=\fdiv{n-1}{n-t}=\fdiv{t+(n-t-1)}{n-t}=\cdiv t{n-t},\]
and $[\theta^n_t]$ is its dual, $\cln{MPT}^k_0$.
\end{Pf}

We will also need a convenient $\tpt$-complete problem to reduce to~$\cmp$. In fact, the statement below effectively
gives a $\tpt$-complete \emph{promise problem} rather than a language. The idea was independently discovered by Wagner
\cite[Cor.~6.4]{wagner:theta2} and Buss and Hay \cite[Thm.~8]{buss-hay}, but the formulation below is closer to
\cite[L.~2.1]{luk-mal}. We include a self-contained proof for completeness.
\begin{Lem}\label{lem:thp2-compl}
Let $L\sset\Sigma^*$ be any language such that $L\in\tpt$. Then there exists a logspace function
$w\mapsto\p{\fii_{w,i}:i<2n_w}$ \paren{where each $\fii_{w,i}$ is a CNF and $n_w\in\N$} with the following property: for every
$w\in\Sigma^*$, there exists $0<j\le2n_w$ such that for all $i<2n_w$,
\[\fii_{w,i}\in\task{Sat}\iff i<j,\]
and we have
\[w\in L\iff j\text{ is even.}\]
\end{Lem}
\begin{Pf}
$L$ is computable by a polynomial-time Turing machine $M(w)$ that makes $\lh w^c$ parallel (non-adaptive) queries to an
$\np$~oracle. Given a $w\in\Sigma^*$, put $n_w=\lh w^c+1$; for any $i<n_w$, let $\fii_{2i}$ be a CNF whose
satisfiability is equivalent to the $\np$ property ``at least $i$ queries made by $M(w)$ have positive answers'',
$\fii'_{2i+1}$ a CNF expressing ``there is an accepting run of $M(w)$ with $i$~positive answers to queries, all of
which are correct'', and $\fii_{2i+1}$ a CNF equivalent to $\fii'_{2i+1}\lor\fii_{2i+2}$ (where $\fii_{2n_w}$ is
understood as~$\bot$). If $k$ is the true number of positively answered queries made by $M(w)$, then $\fii_i$ is
satisfiable iff $i<2k+1$ or $i<2k+2$ depending on whether $w\in L$.
\end{Pf}

The following crucial lemma shows how to leverage threshold functions from Lemma~\ref{lem:thr} to translate the
characterization of~$\tpt$ from Lemma~\ref{lem:thp2-compl} into properties of clones. We recall that threshold functions
have logspace-computable polynomial-size circuits, and even $O(\log n)$-depth formulas (see e.g.\
Wegener~\cite{weg:bool}).
\begin{Lem}\label{lem:th-seq}
There exists a logspace function $\p{\fii_i:i<n}\mapsto f_{\vec\fii}$, and for each $n$, a sequence of integers
$k_{n,0}>k_{n,1}>\dots>k_{n,n}\ge2$, with the following property: whenever $\p{\fii_i:i<n}$ is a sequence of formulas,
we have $[\nto,f_{\vec\fii}]=\cln T^{k_{n,s}}_0$, where $s=\bigl|\{i<n:\fii_i\in\task{Sat}\}\bigr|$.
\end{Lem}
\begin{Pf}
For a given~$n$, fix $m>n$ (to be specified later) and $t=m-n-1$. We may assume that the formulas $\fii_i$ use pairwise
disjoint sets of variables that are also disjoint from $\{x_i:i<m\}$. Put
\[f_{\vec \fii}=\theta^m_t(x_0\land\fii_0,\dots,x_{n-1}\land\fii_{n-1},x_n,\dots,x_{m-1}).\]

When $\fii_i$ is unsatisfiable, we have $x_i\land\fii_i\equiv0$. Thus, renumbering w.l.o.g.\ the $\fii_i$s so that each
$\fii_i$, $i<s$, is satisfiable,
\[f_{\vec\fii}\equiv\theta^{m-n+s}_t(x_0\land\fii_0,\dots,x_{s-1}\land\fii_{s-1},x_n,\dots,x_{m-1}).\]

On the one hand, $x_i\land\fii_i\in\cln T_0^\infty=[\nto]$ by Lemma~\ref{lem:tma}, thus
$f_{\vec\fii}\in[\nto,\theta^{m-n+s}_t]$. On the other hand, for each $i<s$, we may choose a satisfying assignment
$a_i$ to~$\fii_i$, substitute $0\in[\nto]$ for each variable made $0$ by~$a_i$, and substitute $x_i$ for each variable
made $1$ by~$a_i$. (By our assumptions on variables, we can do this independently for each $i<s$, and it will not
affect the $\vec x$ variables.) Under this substitution, $x_i\land\fii_i$ becomes equivalent to~$x_i$, and $f_{\vec\fii}$ becomes
$\theta^{m-n+s}_t(x_0,\dots,x_{s-1},x_n,\dots,x_{m-1})$. Thus,
\[[\nto,f_{\vec\fii}]=[\nto,\theta^{m-n+s}_t]=\cln T_0^{k_{n,s}}\]
using Lemma~\ref{lem:thr}, where
\[k_{n,s}=\cdiv t{m-n-t+s}=\cdiv t{s+1},\]
as long as $m/2+1\le t$, i.e., $m\ge2n+4$. In order to satisfy the constraint $k_{n,0}>k_{n,1}>\dots>k_{n,n}$, it
suffices to ensure that
\[\frac t{s+1}+1\le\frac ts\]
for all $s\le n$, i.e., $t\ge n(n+1)$. This holds if we choose $m=\max\{(n+1)^2,6\}$.
\end{Pf}

We are ready to prove the main result of this section.
\begin{Thm}\label{thm:cmp-main}
$\cmp$ is $\tpt$-complete.
\end{Thm}
\begin{Pf}
We have $\cmp\in\tpt$ by Theorem~\ref{thm:ub}. In order to show that it is $\tpt$-hard, fix $L\in\tpt$.

Given $w$, compute $\p{\fii_i:i<2n_w}$ as in Lemma~\ref{lem:thp2-compl}, and then (abbreviating $n=n_w$)
$f_{\mathrm{even}}=f_{\fii_0,\fii_2,\dots,\fii_{2n-2}}$, $f_{\mathrm{odd}}=f_{\fii_1,\fii_3,\dots\fii_{2n-1}}$ as in
Lemma~\ref{lem:th-seq}. If $j\le2n$ is as in Lemma~\ref{lem:thp2-compl}, then
\begin{align*}
\bigl|\{i<n:\fii_{2i}\in\task{Sat}\}\bigr|&=\cl{j/2},\\
\bigl|\{i<n:\fii_{2i+1}\in\task{Sat}\}\bigr|&=\fl{j/2},
\end{align*}
thus
\begin{align*}
[\nto,f_{\mathrm{even}}]&=\cln T_0^{k_{n,\cl{j/2}}},\\
[\nto,f_{\mathrm{odd}}]&=\cln T_0^{k_{n,\fl{j/2}}},
\end{align*}
where $k_{n,0}>\dots>k_{n,n}\ge2$ are as in Lemma~\ref{lem:th-seq}. It follows that
\[f_{\mathrm{even}}\in[\nto,f_{\mathrm{odd}}]\iff\fl{j/2}=\cl{j/2}\iff j\text{ is even}\iff w\in L.\]
Thus, $w\mapsto\p{\{\nto,f_{\mathrm{odd}}\},f_{\mathrm{even}}}$ is a reduction from $L$ to~$\cmp$.
\end{Pf}

\subsection{Alternative setups}\label{sec:alternative-setups}
%\begin{Rem}\label{rem:nullary}
We followed the tradition in the study of Boolean clones---going back to Post---of considering only functions of
positive arity, even though the general theory of clones and coclones works more smoothly if nullary functions are also
allowed (cf.\ Behrisch~\cite{behr:null}). Let us see now what changes if we take nullary functions into consideration.

First, the number of Boolean clones increases---namely, each non-nullary clone $C$ that includes at least one constant
function (i.e., $C\Sset\cln{UP}_0$ or~$\cln{UP}_1$) splits into two: one consisting only of non-nullary functions as
before, and one that also includes nullary functions corresponding to all constant functions of~$C$. In
Fact~\ref{fact:meet-irr}, we understand the given definitions of meet-irreducible clones so that they include all
applicable nullary functions; moreover, there is a new meet-irreducible clone $\cln N=[\land,\neg]=\pol(r_N)$ of all
non-nullary functions, where $r_N=\nul\in\Rel_1$. Consequently, we include $r_N$ in $R_n$ for each~$n$. Note that
$\cln D\sset\cln N$ and $\cln P\sset\cln N$.

Since the set $\{\land,\lor,\neg\}$ is no longer functionally complete, we read the definition of $\cmp$ and derived
problems so that the input is given in the form of circuits over the basis $\{\land,\lor,\neg,0,1\}$, where $0$ and~$1$
denote nullary constants.

The upper bounds in Theorem~\ref{thm:ub} continue to hold unchanged.

In Theorem~\ref{thm:cmp-C}, the main dichotomy still holds: $\cmp_C\in\ptime$ if $C\Sset\cln P$, and $\cmp_C$ is
$\conp$-complete otherwise. The difference is that now there are more clones $C\Sset\cln P$, namely $\cln\top$, $\cln
N$, $\cln P_0$, $\cln{NP}_0$, $\cln P_1$, $\cln{NP}_1$, and~$\cln P$. $\cmp_\cln\top$ is trivial, and $\cmp_C$ is
$\ptime$-complete for $C=\cln P_\alpha,\cln{NP}_\alpha,\cln P$. The problem $\cmp_\cln N$ amounts to testing if the
given Boolean circuit has a nonzero number of input variables: the exact mechanics of this test will depend on
syntactic details of the representation of input, but it can be done in~$\Ac$ under any reasonable representation.

The statement of Theorem~\ref{thm:cmp-f} changes so that $\cmp^f$ is $\np$-complete if $f$ is neither a projection nor a
nullary function. If $f$ is a nullary function, then $\cmp^f$ is $\ptime$-complete: if $\alpha\in\two$ is a nullary
constant, we have that $\alpha\in[F]$ iff either $\alpha\in F$, or $F$ contains the dual constant $\neg\alpha$ and
$F\nsset\cln P_{\neg\alpha}$ (the second disjunct is only possible if $\lh F\ge2$, thus given a variable-free
circuit~$C$, we have $C=\alpha$ iff $\{C\}\in\cmp^\alpha$, showing the $\ptime$-hardness of $\cmp^f$).

All named clones in Lemmas \ref{lem:thr} and~\ref{lem:th-seq} need to be intersected with~$\cln N$, so that, e.g., the conclusion of
Lemma~\ref{lem:th-seq} reads $[\nto,f_{\vec\fii}]=\cln{NT}_0^{k_{n,s}}$.

The main Theorem~\ref{thm:cmp-main} still holds.
%\end{Rem}

%\begin{Rem}\label{rem:formulas}
A different kind of variation of our results concerns the input representation. We defined $\cmp$ and friends so that
the input functions are represented by Boolean circuits, which is the natural thing to do in a computational context.
However, in the context of logic or algebra, it is more natural to represent Boolean functions by Boolean formulas, or
equivalently, Boolean terms.

Fortunately, this has a negligible effect on our results. First, all upper bounds hold also for the formula
representation, because formulas are special cases of circuits. On the other hand, our main lower bounds continue to
hold in this setting as well: we used reductions from Boolean satisfiability (that already works with formulas), and
the most complicated tools we employed were threshold functions, which can be written with polynomial-size formulas
just as well as circuits.

The only exceptions are problems that were proved $\ptime$-complete by reductions from evaluation of Boolean circuits:
namely, $\cmp_C$ for $\cln P\sset C\ne\cln\top,\cln N$ (Theorem~\ref{thm:cmp-C}), and $\cmp^f$ for $f$ a nullary function
(Theorem~\ref{thm:cmp-f} as modified in the first part of this subsection). If we change the input representation to
formulas, then all these problems are computable in~$\nci$.
%\end{Rem}

\section{Restricted input bases}\label{sec:restr-input-bases}

The problems $\cmp$, $\cmp_C$, and $\cmp^f$ are defined so that the input functions are given by circuits over a
fixed functionally complete basis. This is reasonable if we consider these circuits to be just a computing device.
However, if we view the problem as ``given a circuit over the De~Morgan basis, can we rewrite it as a circuit over a
given basis~$F$?'', it makes perfect sense to also consider the case where the input circuits are expressed in a
different basis. That is, for any finite $B\sset\Op$, let $B$-$\cmp$ be the following problem:
\begin{quote}
{\bf Input:} A finite set of functions $F\sset\Op$ and a function $f\in\Op$, all given by circuits over the basis~$B$.

{\bf Output:} YES if $f\in[F]$, otherwise NO.
\end{quote}
Similarly, we define the problems $B$-$\cmp_C$ and $B$-$\cmp^f$.

The complexity of $B$-$\cmp_C$ was thoroughly investigated by B\"ohler and Schnoor~\cite{boeh-schn}, who denote the
problem as $\mathcal M_C(C\twoheadleftarrow B)$, and its formula version as $\mathcal M(C\twoheadleftarrow B)$. For
most combinations of $C$ and~$B$, they were able to show that $\mathcal M(C\twoheadleftarrow B)$ and $\mathcal
M_C(C\twoheadleftarrow B)$ are both $\conp$-complete, or both in~$\ptime$.

We will not make any effort to classify the complexity of the problems $B$-$\cmp^f$, as the number of cases is
prohibitively large, hence we consider it out of scope of this paper. (The problem has two clone parameters,
analogously to $B$-$\cmp_C$, which took B\"ohler and Schnoor a whole paper to understand, and even then their
classification is incomplete.)

We will not obtain a complete classification of the complexity of $B$-$\cmp$ either, nevertheless we present a number of
partial results, summarized in Corollary~\ref{cor:b-cmp-summ} at the end of the section.

Unless stated otherwise, the complexity results below also hold for variants of $B$-$\cmp$ where the input
functions are represented by formulas (in prefix or infix notation).
For simplicity, we disallow nullary functions in this section, but the results below can be easily adapted to a setup
that includes them. Our results will be stated for arbitrary bases, but we will often need to express specific
functions: to this end, we recall that we can more-or-less freely convert circuits and formulas to different bases by
Lemma~\ref{lem:conv}. In particular, Lemma~\ref{lem:conv}~\ref{item:8} applies to polynomial-size CNFs and other
constant-depth formulas using unbounded fan-in $\land$, $\lor$ gates, as these can be written as $O(\log n)$-depth
bounded fan-in formulas.

For the rest of this section, $B$ is a finite subset of~$\Op$.

We start with a simple observation.
\begin{Lem}\label{lem:b-cmp-bh}
If $B\sset\cln T_0^\infty$, $\cln T_1^\infty$, $\cln\ET$, $\cln\LOR$, $\cln A$, or~$\cln D$, then
$B\text-\cmp\in\cxt{BH}$.
\end{Lem}
\begin{Pf}
By inspection of Post's lattice, we see that there are only finitely many clones below~$[B]$. Thus, if $F$ and~$f$ are
given by $B$-circuits, we have
\[f\in[F]\iff\forall r\in R_k\,(F\npres r\text{ or }f\pres r)\]
for some constant~$k$, which gives a $\cxt{BH}$ algorithm in view of Lemma~\ref{lem:pres-conp}.
\end{Pf}

This suggests that there is a major difference between the complexity of $B$-$\cmp$ in cases where $[B]$ has finitely
many subclones and in the other cases. We will study the former in Section~\ref{sec:finite-subclone-case}, and the
latter in Section~\ref{sec:hard-cases}.

\subsection{The finite subclone cases}\label{sec:finite-subclone-case}

We are going to characterize more precisely the complexity of $B$-$\cmp$ for $B$ as in Lemma~\ref{lem:b-cmp-bh}, i.e.,
such that $[B]$ has finitely many subclones. Depending on~$[B]$, we will classify it as being either
$\cxt{coDP}$-complete or in~$\ptime$ (with the formula version in~$\nci$). We are first going to deal with the $\ptime$
cases, then show $\cxt{coDP}$-hardness of the remaining cases using reductions from variants of the equivalence
problem, and finish by establishing they are indeed in~$\cxt{coDP}$.

We start with the tractable cases. The theorem below mostly follows from results of B\"ohler and
Schnoor~\cite{boeh-schn}: since $[B]$ has only finitely many subclones, we have $B$-$\cmp\in\ptime$ iff
$B$-$\cmp_C\in\ptime$ for all clones $C\sset[B]$. However, we give a self-contained proof, which also confirms that the
complexity drops down to~$\nci$ for formulas; the basic idea is that the clones in question have so simple structure
that we can determine $[f]$ by evaluating $f$ on a specific polynomial-size set of assignments.
\begin{Thm}\label{thm:b-cmp-ptime}
If $B\sset\cln{MT}^\infty_0$, $\cln{MT}^\infty_1$, $\cln\ET$, $\cln\LOR$, $\cln A$,
or~$\cln{DM}$, then $B\text-\cmp\in\ptime$. If we represent the input by formulas rather than circuits,
$B\text-\cmp\in\nci$.
\end{Thm}
\begin{Pf}
Assume that $B\sset\cln\LOR$. The set
of $n$-ary functions in $\cln\LOR$ is very limited: it consists only of $1$ and the functions
$f_I(x_0,\dots,x_{n-1})=\LOR_{i\in I}x_i$ for $I\sset\{0,\dots,n-1\}$ (including $f_\nul(\vec x)=0$). Given a $B$-circuit, we
can determine which of these functions it computes by evaluating it on the assignment $\vec0$, which detects the $1$
function, and for each $i<n$, on $e_i=\p{0,\dots,0,1,0,\dots,0}$ (with $1$ at position~$i$), which detects if $i\in I$.

Once we know the function, it is trivial to determine which of the finitely many subclones of~$\cln\LOR$ it generates,
and if we know the clones generated by $f$ and by each element of~$F$, we can find out if $f\in[F]$. This gives a
polynomial-time algorithm.
Moreover, since the algorithm just evaluates the input functions on polynomially many assignments
in parallel, and then does $\Ac$ post-processing, it can be implemented in~$\nci$ if the functions are given by
formulas rather than circuits.

The cases of $B\sset\cln\ET$ or $B\sset\cln A$ are completely analogous to $\cln\LOR$. In particular, the $n$-ary
functions in~$\cln A$ are $f_{\vec\alpha,\alpha}(\vec x)=\sum_{i<n}\alpha_ix_i+\beta$ for some
$\alpha_0,\dots,\alpha_{n-1},\beta\in\two$; given a $B$-circuit computing a function~$f$, we can determine $\vec\alpha$
and~$\beta$ such that $f=f_{\vec\alpha,\beta}$ by evaluating $\beta=f(\vec0)$ and $\alpha_i=f(e_i)+\beta$.

Assume that $[B]=\cln{DM}$. Since $\cln{DM}$ is a minimal clone, a $B$-circuit either computes a projection, or it
generates~$\cln{DM}$. Moreover, a self-dual monotone function $f\in\Op_n$ is the projection $\pi^n_i$ iff
$f(e_i)=1$. Thus, we can again determine $[f]$ by evaluating $f$ at $n$~assignments.

Assume that $B\sset\cln{MT}^\infty_1$, the case of $\cln{MT}^\infty_0$ being dual. Given a $B$-circuit computing a
function~$f$, either $f\in\cln\LOR$, or $[f]=\cln{MPT}^\infty_1$, $\cln{MT}^\infty_1$. By evaluating $f(e_i)$
for all~$i<n$, we find the only candidate $I\sset\{0,\dots,n-1\}$ such that $f$ could equal $f_I$. Then we evaluate $f$
at the assignment $a$ such that $a_i=1$ iff $i\notin I$: if $f(a)=0$, then $f\equiv f_I$, otherwise $f\notin\cln\LOR$.
In the latter case, we can distinguish $\cln{MT}^\infty_1$ from~$\cln{MPT}^\infty_1$ by evaluating $f(\vec0)$.
\end{Pf}

We remark that while $B$-$\cmp$ is $\ptime$-complete (or $\nci$-complete in the formula representation) for
$[B]=\cln{M(P)T}^\infty_\alpha$ or $[B]=\cln{DM}$, it is still easier for other classes from Theorem~\ref{thm:b-cmp-ptime}:
for example, it is easy to show that if $[B]\sset\cln\LOR$, we can evaluate $B$-circuits in~$\cxt L$, and $B$-formulas
in~$\Ac$. We will not go into details.

We now turn to the remaining clones below $\cln T^\infty_\alpha$ and $\cln D$. We will use the following result for lower
bounds. Here, the \emph{equivalence problem} $B$-$\Eq$ is to compute if two given Boolean formulas over basis~$B$ are
equivalent.
\begin{Thm}[Reith~\cite{reith:phd}]\label{thm:eq-conp}
If $[B]\Sset\cln{MPT}^\infty_0$, $\cln{MPT}^\infty_1$, or $\cln{DM}$, then $B$-$\Eq$ is $\conp$-complete.
\noproof\end{Thm}
We mention that since the formulas used in Theorem~\ref{thm:eq-conp} are built from DNFs, they can be taken to
have depth $O(\log n)$. In particular, this ensures they can be efficiently converted to a different basis by
Lemma~\ref{lem:conv}.

The constructions in the next lemma mostly come from B\"ohler and Schnoor~\cite{boeh-schn}, who used them to prove
$\conp$-completeness of various instances of $B$-$\cmp_C$. We observe that the lower bound can be improved to
$\cxt{coDP}$ if $C$ is allowed to vary.
\begin{Lem}\label{lem:b-cmp-codp-hard}
If $[B]\Sset\cln{PT}^\infty_0$, $\cln{PT}^\infty_1$, $\cln{MPT}^2_0$, $\cln{MPT}^2_1$, or $\cln{DP}$, then
$B$-$\cmp$ is $\cxt{coDP}$-hard.
\end{Lem}
\begin{Pf}
Assume first $[B]\Sset\cln{DP}$. If $f,g$ are $\cln{DM}$-formulas (i.e., formulas over a fixed basis of~$\cln{DM}$),
let $h_{f,g}(\vec x,y_0,y_1,y_2)=f(\vec x)+g(\vec x)+\theta^3_2(\vec y)$; this can be expressed by a $\cln{DP}$-formula
as the ternary function $x+y+z$ is in~$\cln{DP}$. In view of $f,g\in\cln P$, we have
$\theta^3_2(\vec y)=h_{f,g}(x,\dots,x,\vec y)\in[h_{f,g}]$, thus $\cln{DM}\sset[h_{f,g}]$. If $f\equiv g$, then
$h_{f,g}(\vec x,\vec y)\equiv\theta^3_2(\vec y)$. If $f\nequiv g$, let $\vec a$ be an assignment such that
$f(\vec a)\ne g(\vec a)$. Then $h_{f,g}(\vec0,\vec1)=1$ (using $f,g\in\cln P$) and $h_{f,g}(\vec a,\vec1)=0$, hence
$h_{f,g}$ is not monotone. Since $\cln{DM}\sset[h_{f,g}]\sset\cln{DP}$, this implies $[h_{f,g}]=\cln{DP}$. Thus,
\begin{align}
\label{eq:4}f\equiv g&\implies[h_{f,g}]=\cln{DM},\\
\label{eq:5}f\nequiv g&\implies[h_{f,g}]=\cln{DP}.
\end{align}
Now, since $\cln{DM}$-$\Eq$ is $\conp$-complete by Theorem~\ref{thm:eq-conp}, the language
\[L=\bigl\{\p{f,g,f',g'}:\p{f,g}\in\cln{DM}\text-\Eq\text{ or }\p{f',g'}\notin\cln{DM}\text-\Eq\bigr\}\]
is $\cxt{coDP}$-complete. Using \eqref{eq:4} and~\eqref{eq:5},
\[\p{f,g,f',g'}\in L\iff[h_{f,g}]\sset[h_{f',g'}],\]
which gives a reduction of $L$ to $B$-$\cmp$.

If $[B]\Sset\cln{MPT}^2_1$ (the case of $\cln{MPT}^2_0$ is dual), we can use in a similar way the
$\cln{MPT}^2_1$-formula
\[h(\vec x,y,z)=\theta^3_2\bigl(f(\vec x)\lor g(\vec x),y,z\bigr).\]
The dual of $h$ is $\theta^3_2\bigl(f(\vec x)\land g(\vec x),y,z\bigr)$, hence $h$ is self-dual if and only if $f\equiv
g$. Moreover, $f(x,\dots,x)\equiv g(x,\dots,x)\equiv x$, hence in any case $\theta^3_2\in[h]$. Thus,
\begin{align*}
f\equiv g&\implies[h]=\cln{DM},\\
f\nequiv g&\implies[h]=\cln{MPT}^2_1.
\end{align*}
This gives a reduction of $L$ to $B$-$\cmp$ in the same way as above.

Finally, assume $[B]\Sset\cln{PT}^\infty_1$ (the case of $\cln{PT}^\infty_0$ is dual). Given
$\cln{MPT}^\infty_1$-formulas $f$ and~$g$, we put $h(\vec x,y)=y\lor\bigl(f(\vec x)+g(\vec x)\bigr)$, which can be
expressed by a $\cln{PT}^\infty_1$-formula. If $f\equiv g$, then $h\equiv y$; otherwise, we check easily that $h$ is
not monotone. Thus,
\begin{align*}
f\equiv g&\implies[h]=\bot,\\
f\nequiv g&\implies[h]=\cln{PT}^\infty_1,
\end{align*}
which yields a reduction of the $\cxt{coDP}$-complete problem
\[\bigl\{\p{f,g,f',g'}:\p{f,g}\in\cln{MPT}^\infty_1\text-\Eq\text{ or }\p{f',g'}\notin\cln{MPT}^\infty_1\text-\Eq\bigr\}\]
to $B$-$\cmp$.
\end{Pf}

Incidentally, the argument we gave for $\cln{MPT}^2_\alpha$ also resolves one of the problems left open by B\"ohler and
Schnoor~\cite{boeh-schn}:
\begin{Cor}\label{cor:dm-conp}
If $[B]\Sset\cln{MPT}^2_0$ or $\cln{MPT}^2_1$, then $B$-$\cmp_\cln{DM}$ is $\conp$-complete.
\noproof\end{Cor}

It remains to prove $\cxt{coDP}$ upper bounds for the $B$-$\cmp$ problems in question. We obtain this by optimizing the
number of $\np$~oracle calls in the algorithm from Lemma~\ref{lem:b-cmp-bh}, exploiting the fact that we can distinguish
between most proper subclones of~$[B]$ in polynomial time by Theorem~\ref{thm:b-cmp-ptime}.

\begin{Thm}\label{thm:codp-compl}
If $[B]$ is $\cln T^\infty_0$, $\cln{PT}^\infty_0$, $\cln T^\infty_1$, $\cln{PT}^\infty_1$, $\cln D$, or $\cln{DP}$,
then $B$-$\cmp$ is $\cxt{coDP}$-complete.
\end{Thm}
\begin{Pf}
$\cxt{coDP}$-hardness was proved in Lemma~\ref{lem:b-cmp-codp-hard}. Assume that $B\sset\cln T^\infty_1$, we will refine
Lemma~\ref{lem:b-cmp-bh} to show that $B\text-\cmp\in\cxt{coDP}$.

First, given a set of $B$-circuits $F$, we can determine $[F]$ in polynomial time using a single $\conp$ oracle query,
namely $F\overset?\sset\cln M$: indeed, if $F\nsset\cln M$, then $[F]$ is $\cln{PT}^\infty_1$ or $\cln T^\infty_1$, and
we can distinguish these two cases by testing if $F\sset\cln P$ (or equivalently, $\cln P_0$), which we can do by
evaluating $g(\vec0)$ for each $g\in F$. On the other hand, if $F\sset\cln M$, i.e., $F\sset\cln{MT}^\infty_1$, we can
compute $[F]$ using the algorithm in Theorem~\ref{thm:b-cmp-ptime} (which again proceeds by evaluation of $F$ at various
assignments, hence it does not matter that the circuits are given in a larger basis).

This already shows that $B\text-\cmp\in\ptime^{\|\np[2]}$: we can test if $f\in[F]$ using two parallel $\conp$
queries, $F\overset?\sset\cln M$ and $f\overset?\in\cln M$.

In order to improve this to $\cxt{coDP}$, we modify the algorithm so that it speculatively explores all computation
branches with all possible oracle answers, and only makes the oracle queries needed at the end.

In this way, the algorithm computes two candidate clones $C_0\sset\cln{MT}^\infty_1$ and $C_1\Sset\cln{PT}^\infty_1$
for~$[F]$. We have $C_0\sset C_1$: the choice of $C_1$ as $\cln{PT}^\infty_1$ or~$\cln T^\infty_1$ is
made according to if $F\sset\cln P$, and this information is taken into account also when computing~$C_0$. In fact,
this means that $C_1$ is the join $C_0\lor\cln{PT}^\infty_1$ in the lattice of clones.

(Actually, the algorithm may fail to compute $C_0$ because it runs into an inconsistency that already shows
$F\nsset\cln M$. In this case, we may pick $C_0$ in an arbitrary way such that the properties $C_0\sset\cln M$ and
$C_1=C_0\lor\cln{PT}^\infty_1$ hold, say, $C_0=\bot$ or $C_0=\cln{UP}_1$: if the choice of~$C_0$ became relevant in
subsequent computation, it would be dismissed by the oracle as $F\nsset\cln M$.)
 
Likewise, we obtain two candidate clones $C'_0$, $C'_1$ for~$[f]$, with $C'_0\sset\cln{MT}^\infty_1$ and
$C'_1=C'_0\lor\cln{PT}^\infty_1$.

Notice that $C'_1\nsset C_0$ as $C_0\sset\cln M$, and that $C'_0\sset C_1$ implies $C'_1=C'_0\lor\cln{PT}^\infty_1\sset
C_1$. Thus, there are only the following possibilities:
\begin{itemize}
\item $C'_0\sset C_0$ (whence $C'_1\sset C_1$): then $f\in[F]$ if and only if $F\nsset\cln M$ or $f\in\cln M$.
\item $C'_1\sset C_1$ and $C'_0\nsset C_0$: then $f\in[F]$ if and only if $F\nsset\cln M$.
\item $C'_0\nsset C_1$: then $f\notin[F]$.
\end{itemize}
Consequently, the whole algorithm can be implemented in $\cxt{coDP}$: we have
\[f\in[F]\iff\p{F,f}\in L_0\text{ or }\p{F,f}\in L_1\]
with
\begin{align*}
L_0&=\bigl\{\p{F,f}:C'_1\sset C_1\text{ and }F\nsset\cln M\bigr\}\in\np,\\
L_1&=\bigl\{\p{F,f}:C'_0\sset C_0\text{ and }f\in\cln M\bigr\}\in\conp,
\end{align*}
where $C_0$, $C_1$, $C'_0$, $C'_1$ are computed from $\p{F,f}$ in deterministic polynomial time as described above.

The case of $B\sset\cln T^\infty_0$ is dual.

The argument for $B\sset\cln D$ uses the same overall strategy: given $F$, we compute in polynomial time two candidates
$C_0$ and $C_1=C_0\lor\cln{DP}$ for~$[F]$, where $C_0\sset\cln{DM}$ or $C_0\sset\cln{AD}$. Thus, $[F]$ is
$C_0$ if $F\sset\cln M$ or $F\sset\cln A$, and it is~$C_1$ otherwise. Likewise, given~$f$, we compute candidates
$C'_0$ and $C'_1=C'_0\lor\cln{DP}$ for~$[f]$ satisfying $C'_0\sset\cln{DM}$ or $C'_0\sset\cln{AD}$. Then as above,
$f\in[F]$ iff $\p{F,f}\in L_0\text{ or }\p{F,f}\in L_1$, with
\begin{align*}
L_0&=\bigl\{\p{F,f}:C'_1\sset C_1\text{ and }F\nsset\cln M\text{ and }F\nsset\cln A\bigr\}\in\np,\\
L_1&=\bigl\{\p{F,f}:C'_0\sset C_0\text{ and }(f\in\cln M\text{ or }f\in\cln A)\bigr\}\in\conp.
\end{align*}

The candidates are determined as follows. If $g(\vec0)=1$ for some $g\in F$, we put $C_1=\cln D$, and $C_0$ is
$\cln{AD}$ or $\cln{UD}$ depending on whether there exists $g\in F$ such that $\lh{\{i:g(e_i)\ne g(\vec0)\}}>1$.

If $g(\vec0)=0$ for all $g\in F$, then $C_1=\cln{DP}$. Putting $m(g)=\lh{\{i:g(e_i)=1\}}$, the argument in the proof of
Theorem~\ref{thm:b-cmp-ptime} shows that if $F\sset\cln{DM}$, then $m(g)\in\{0,1\}$ for all $g\in F$, whereas if
$F\sset\cln{AP}$, then $m(g)$ is odd for all $g\in F$; in both cases, $m(g)=1$ iff $g$ is a projection. Thus, we may
define $C_0=\cln{DM}$ if $m(g)=0$ for some $g\in F$, $C_0=\cln{AP}$ if $m(g)>1$ for some $g\in F$, and $C_0=\bot$
otherwise.
\end{Pf}

\subsection{The hard cases}\label{sec:hard-cases}

The question we are mainly interested in is for which bases~$B$ is $B$-$\cmp$ a $\tpt$-complete problem. We first adapt
Theorem~\ref{thm:cmp-main} easily to $\cln P$-$\cmp$ using simple transformations to eliminate constants. Next, we
generalize it further to the clones $\cln{PT}^k_\alpha$ by expressing the threshold functions used in the
proof of Lemma~\ref{lem:th-seq} in a $\{\theta^{k+1}_k\}$ basis; we only manage to do so with a randomized construction,
leading to a $\tpt$-completeness result under a suitable notion of randomized reductions.

In order to get the result for $\cln P$-$\cmp$, we use the following translation (for completeness, we formulate it
more generally than what we need):
\begin{Lem}\label{lem:transl-p}
Let $B,B'\sset\Op$ be finite. Assume that $\land\in[B']$ and $B\sset[B',0]$, or $\lor\in[B']$ and $B\sset[B',1]$, or
$\land,\lor\in[B']$ and $B\sset[B',0,1]$.
\begin{enumerate}
\item\label{item:9}
Given a $B$-circuit that computes a $B'$-function~$f$, we can construct in~logspace a $B'$-circuit that computes~$f$.
\item\label{item:10}
Given a $B$-formula of depth $O(\log n)$ that computes a $B'$-function~$f$, we can construct in logspace a $B'$-formula
that computes~$f$.
\end{enumerate}
In particular, this applies to arbitrary $B$ if $[B']\Sset\cln P$.
\end{Lem}
\begin{Pf}
Assume first $\land\in[B']$ and $B\sset[B',0]$. Without loss of generality, $0\notin[B']$, hence $B'\sset\cln P_1$.

Since $B\sset[B',0]$, Lemma~\ref{lem:conv} implies that we can find a $B'$-circuit~$g$ (or even an $O(\log n)$-depth
$B'$-formula) such that $f(\vec x)\equiv g(\vec x,0)$. Then $f(\vec x)\equiv g\bigl(\vec x,\ET_ix_i\bigr)$:
the two expressions agree when $\vec x\ne\vec1$ as $\ET_ix_i=0$; they also agree for
$\vec x=\vec1$ as $f(\vec1)=1=g(\vec1,1)$ on account of $f,g\in\cln P_1$. Since
$\ET_ix_i$ has an $O(\log n)$-depth formula over the basis $\{\land\}\sset[B']$, it can be
written by an $O(\log n)$-depth $B'$-formula using Lemma~\ref{lem:conv} again.

The case $\lor\in[B']$ and $B\sset[B',1]$ is dual.

Let $\land,\lor\in[B']$ and $B\sset[B',0,1]$. If $0\in[B']$ or $1\in[B']$, we are done by one of the previous cases,
hence we may assume $0,1\notin[B']$, which implies $B'\sset\cln P$.
Then we proceed as before: we find a $B'$-circuit $g$ such that $f(\vec x)\equiv g(\vec x,0,1)$, and we observe that
$f(\vec x)\equiv g\bigl(\vec x,\ET_ix_i,\LOR_ix_i\bigr)$ because of $f,g\in\cln P$.
\end{Pf}

\begin{Thm}\label{thm:p-cmp-th2}
If $[B]\Sset\cln P$, then $B$-$\cmp$ is $\tpt$-complete.
\end{Thm}
\begin{Pf}
The formulas we constructed in the proof of Theorem~\ref{thm:cmp-main} (or rather, Lemma~\ref{lem:th-seq}) compute functions
in~$\cln P_0$, and have depth $O(\log n)$ (being build from CNFs and threshold functions), hence we can convert them to
$\cln P_0$-formulas by Lemma~\ref{lem:transl-p}. Thus, in the setting of Lemma~\ref{lem:th-seq}, we map a sequence of
formulas $\p{\fii_i:i<n}$ to a $\cln P_0$-formula $f_{\vec\fii}(\vec x)$ such that $[\nto,f_{\vec\fii}]=\cln T^k_0$,
where $k\ge2$ is as specified in the lemma.

Using the idea of Lemma~\ref{lem:transl-p}, we can construct a $\cln P$-formula $g_{\vec\fii}(\vec x,y)$ (and therefore a
$B$-formula, using Lemma~\ref{lem:conv}) such that $f_{\vec\fii}(\vec x)\equiv g_{\vec\fii}(\vec x,0)$. By replacing
$g_{\vec\fii}$ with $g_{\vec\fii}\bigl(\vec x,y\land\ET_ix_i\bigr)$ if necessary, we ensure that
\begin{equation}\label{eq:7}
g_{\vec\fii}(\vec x,y)\equiv f_{\vec\fii}(\vec x)\lor\Bigl(y\land\ET_ix_i\Bigr).
\end{equation}
We claim that
\begin{equation}\label{eq:6}
[x\land(y\to z),g_{\vec\fii}]=\cln{PT}^k_0,
\end{equation}
hence we can use $[x\land(y\to z),g_{\vec\fii}]$ in place of $[\nto,f_{\vec\fii}]$ in the proof of
Theorem~\ref{thm:cmp-main} to get a reduction from any $\tpt$~language to $B$-$\cmp$.

In order to prove~\eqref{eq:6}, observe that $[x\land(y\to z)]=\cln{PT}^\infty_0$: clearly $x\land(y\to z)\in\cln P$,
and $x\land(y\to z)\in\cln T^\infty_0$ by Lemma~\ref{lem:tma}, while $x\land(y\to z)\notin\cln M$. Thus,
$\cln{PT}^\infty_0\sset[x\land(y\to z),g_{\vec\fii}]\sset\cln P$
and
$[x\land(y\to z),g_{\vec\fii},0]\Sset[\nto,f_{\vec\fii}]=\cln T^k_0$,
which implies
\[[x\land(y\to z),g_{\vec\fii}]=\cln{PT}^l_0\]
for some $1\le l\le k$. It now suffices to show that $g_{\vec\fii}\in\cln T^k_0=\pol(r^k_0)$.

Assume for contradiction that there are assignments $\vec a^0$, \dots, $\vec a^{k-1}$ such that $\vec
a^0\land\dots\land\vec a^{k-1}=\vec0$, but $g_{\vec\fii}(\vec a^0)=\dots=g_{\vec\fii}(\vec a^{k-1})=1$. Notice that if
$\vec a^i=\vec1$ for some~$i$, we may leave it out (or rather, replace it with another assignment from
$g_{\vec\fii}^{-1}[1]$), as still $\ET_{j\ne i}\vec a^j=\vec0$. Thus, we may assume that none of the $\vec a^i$ is
$\vec1$. But then $f_{\vec\fii}(\vec a^i)=g_{\vec\fii}(\vec a^i)=1$ for each $i<k$ by~\eqref{eq:7} (more precisely,
this holds for the truncation of $\vec a^i$ leaving out the last coordinate). This contradicts $f_{\vec\fii}\in\cln
T^k_0$.
\end{Pf}

Now we would like to extend Theorem~\ref{thm:p-cmp-th2} to the clones $\cln{PT}^k_\alpha$, and for that we need efficient
constructions of threshold functions in a $\cln{PT}^k_\alpha$~basis. This in fact appears to be quite a challenging
task. The best deterministic result we found is a construction by Cohen et al.~\cite{cdikmrr:thr}, whose special case
for the $\{\theta^{k+1}_2\}$ basis is as follows:
\begin{Thm}[Cohen et al.~\cite{cdikmrr:thr}]\label{thm:thr-form}
Let $k\ge2$. There exists a constant $c$ and a polynomial-time algorithm that, given a sufficiently large~$N$,
constructs in time $N^{O(1)}$ an $O(\log N)$-depth $\{\theta^{k+1}_2\}$-formula $\psi_N(x_0,\dots,x_{N-1})$ such that
\[\theta^N_{\fl{N(k^{-1}+\ep)}}\le\psi_N\le\theta^N_{\cl{N(k^{-1}-\ep)}},\]
where $\ep=c/\sqrt{\log N}$.
\noproof\end{Thm}

Since the formula~$\psi_N$ can only reliably distinguish inputs whose Hamming weights differ by
$\Omega(N/\sqrt{\log N})$, it cannot be used to tell apart more than $O(\sqrt{\log N})$ different cases. Thus, in order
to distinguish $n$ possible outcomes as in Lemma~\ref{lem:th-seq}, we would need $N=\exp(\Omega(n^2))$, which makes it
useless for our purposes. For the special case $k=2$, they give a better construction that reduces the error of
approximation to $2^{-O(\sqrt{\log N})}$, which means we might get away with $N=n^{O(\log n)}$, but this is still
insufficient.

In absence of a better idea, we resort to probabilistic constructions following the method of Valiant~\cite{val:maj},
who used it to prove the existence of short $\{\land,\lor\}$-formulas for majority; Gupta and Mahajan~\cite{gup-mah}
modified his construction to produce $\{\theta^3_2\}$-formulas. We will use a similar idea to express
suitable threshold functions by short formulas over the $\{\theta^{k+1}_k\}$ basis.
\begin{Thm}\label{thm:rand-th-fla}
Let $k\ge3$. There exist constants $\frac12<\sigma_k<1$ and $c\ge1$, and a logspace function~$T$ with the following
properties. The input of $T$ consists of numbers $n$, $t$, and $e$ in unary, and $r\in\{0,1\}^*$; the output is a
$\{\theta^{k+1}_k\}$-formula $T_{n,t,e,r}(x_0,\dots,x_{n-1})$ of depth $O(\log n+\log e)$. If $n$ is sufficiently large
and $\sigma_kn<t\le n$, then
\[\PR_{\lh r=(n+e)^c}[T_{n,t,e,r}\equiv\theta^n_t]\ge1-2^{-e}.\]
\end{Thm}
\begin{Pf}
Given $n$ and $d$, let $F_{n,d}$ be a random formula consisting of a complete $(k+1)$-ary tree of
$\theta^{k+1}_k$-gates of depth~$d$, where each leaf is a propositional variable independently uniformly chosen from
$\{x_i:i<n\}$. If $a\in\two^n$ is an assignment of weight $w=pn$, $p\in[0,1]$, then $F_{n,d}(a)$ is a Bernoulli random
variable that takes value~$1$ with certain probability~$p_d$. We can describe $F_{n,d}(a)$ as the value of a complete
$(k+1)$-ary tree of depth~$d$ of $\theta^{k+1}_k$-gates, where each leaf is an independently drawn random element
of~$\two$ with the probability of $1$ being~$p$ (which also makes it manifest that---as suggested by the
notation---$p_d$ only depends on $p$ and~$d$, not on~$a$). Since the $k+1$ input subformulas of any gate are
independent, this gives a recurrence for~$p_d$:
\begin{align*}
p_0&=p,\\
p_{d+1}&=f(p_d),
\end{align*}
where
\[f(x)=x^{k+1}+(k+1)x^k(1-x)=(k+1)x^k-kx^{k+1}.\]
Clearly, $f$ maps $[0,1]$ to $[0,1]$. In order to analyze the behaviour
of~$p_d$, we need to locate the fixed points of~$f$ in~$[0,1]$, i.e., the roots of the polynomial $g(x)=f(x)-x$.
The end-points $0$ and~$1$ are roots. Moreover, the derivative
\[g'(x)=(k+1)kx^{k-1}(1-x)-1\]
satisfies $g'(0)=g'(1)=-1<0$, hence there must be another root in $(0,1)$. On the other hand, Descartes's
rule of signs implies that $g$ has at most two positive roots. Thus, $g$ has a \emph{unique} root $\sigma_k=\sigma\in(0,1)$;
$g$ is negative on $(0,\sigma)$, and positive on $(\sigma,1)$. That is, $\sigma$ is the unique fixed point of $f$ in
$(0,1)$, and we have
\begin{align*}
0<x<\sigma&\implies f(x)<x,\\
\sigma<x<1&\implies x<f(x).
\end{align*}
Consequently, for any $p$, the sequence $p_d$ is monotone (decreasing for $0<p<\sigma$, and increasing for
$\sigma<p<1$), and as such it has a limit, which must be a fixed point of~$f$. Thus,
\begin{align*}
0\le p<\sigma&\implies\lim_{d\to\infty}p_d=0,\\
\sigma<p\le1&\implies\lim_{d\to\infty}p_d=1.
\end{align*}

We claim that $\sigma$ is irrational, hence $p\ne\sigma$: since $\sigma$ is a root of $1-(k+1)x^{k-1}+kx^k$,
$\sigma^{-1}$ is a root of the monic polynomial $h(x)=x^k-(k+1)x+k$, and as such, it is an algebraic integer. Thus,
if it were rational, it would be an actual integer; however, it is easy to see that $h(x)>0$ for all $x\ge2$, hence
$1<\sigma^{-1}<2$.

Next, we need to analyze the rate of convergence of~$p_d$. In the vicinity of~$\sigma$, we have
\[f(x)=x+(x-\sigma)g'(\sigma)+O\bigl((x-\sigma)^2\bigr),\]
where $g'(\sigma)>0$: since $g(0)=g(\sigma)=g(1)=0$, $g'$ has a root in $(0,\sigma)$ and
another in $(\sigma,1)$, while it has at most two positive roots altogether by the rule of signs, hence
$g'(\sigma)\ne0$. We cannot have $g'(\sigma)<0$ as $g$ is negative on $(0,\sigma)$ and positive on $(\sigma,1)$.

Thus, we may fix constants $\ep_0>0$ and $\gamma_0>1$ such that
\[\Abs{x-\sigma}\le\ep_0\implies\Abs{f(x)-\sigma}\ge\gamma_0\Abs{x-\sigma},\]
hence
\[\Abs{p_d-\sigma}\ge\min\bigl\{\ep_0,\gamma_0^d\Abs{p-\sigma}\bigr\}.\]
By Roth's theorem, $\sigma$ has irrationality measure~$2$, i.e., for any $\delta>0$, all but finitely many pairs of
integers $a,b>0$ satisfy
\[\Abs{\frac ab-\sigma}\ge b^{-(2+\delta)}.\]
(The weaker theorem of Liouville, bounding the irrationality measure by $k-1$, would also work for our purposes.)
Applying this to $p=w/n$, we obtain
\[\log\Abs{p-\sigma}^{-1}\le(2+o(1))\log n,\]
thus there exists a constant~$c_0$ such that
\begin{equation}\label{eq:9}
d\ge c_0\log n\implies\Abs{p_d-\sigma}\ge\ep_0
\end{equation}
for any sufficiently large $n$ and $p=w/n$.

In the vicinity of the end-points, we have
\begin{align}
f(x)&=(k+1)x^k+O(x^{k+1}),\\
\label{eq:8}f(1-x)&=1-\binom{k+1}2x^2+O(x^3).
\end{align}
Thus, we may fix $\ep_1>0$ and $\gamma_1>0$ such that
\[0\le x\le\ep_1\implies f(x)\le\gamma_1x^2\text{ and }f(1-x)\ge1-\gamma_1x^2,\]
hence
\begin{alignat}{3}
\label{eq:12}0\le p_d&\le\ep_1&&\implies&p_{d+d'}&\le\gamma_1^{-1}(\gamma_1p)^{2^{d'}},\\
\label{eq:13}1-\ep_1\le p_d&\le1&&\implies&p_{d+d'}&\ge1-\gamma_1^{-1}\bigl(\gamma_1(1-p)\bigr)^{2^{d'}}.
\end{alignat}
We may assume $\gamma_1\ep_1<1$. There is a constant $d_0$ such that
\begin{equation}\label{eq:10}
\Abs{p_d-\sigma}\ge\ep_0\implies p_{d+d_0}\le\ep_1\text{ or }1-p_{d+d_0}\le\ep_1.
\end{equation}
Putting \eqref{eq:9}, \eqref{eq:12}, \eqref{eq:13} and~\eqref{eq:10} together, there is a constant $c_1$ such that
\[d\ge c_1(\log n+\log e)\implies p_d\le2^{-n-e}\text{ or }1-p_d\le2^{-n-e}\]
for sufficiently large~$n$; that is, for any assignment $a\in\two^n$ of weight~$w$,
\[d\ge c_1(\log n+\log e)\implies\begin{cases}
  \PR[F_{n,d}(a)\ne0]\le2^{-n-e},&w<\sigma n,\\
  \PR[F_{n,d}(a)\ne1]\le2^{-n-e},&w>\sigma n.
\end{cases}\]
Using the union bound over all assignments $a\in\two^n$, we obtain
\[d\ge c_1(\log n+\log e)\implies\PR[F_{n,d}\equiv\theta^n_{\cl{\sigma n}}]\ge1-2^{-e}.\]
Now, given $n$ and $t$ such that $\sigma n<t\le n$, put $N=\fl{\sigma^{-1}t}\ge n$. Then $\cl{\sigma N}=t$, thus
\[\theta^n_t(\vec x)\equiv\theta^N_{\cl{\sigma N}}(\vec x,0,\dots,0)
  \equiv\textstyle\theta^N_{\cl{\sigma N}}\bigl(\vec x,\ET_ix_i,\dots,\ET_ix_i\bigr),\]
and consequently
\[d\ge c_2(\log n+\log e)\implies
      \PR\bigl[{\textstyle F_{N,d}\bigl(\vec x,\ET_ix_i,\dots,\ET_ix_i\bigr)}\equiv\theta^n_t\bigr]\ge1-2^{-e}\]
for some constant~$c_2$. We thus define
\[T_{n,t,e,r}=F_{\fl{\sigma^{-1}t},c_2(\log n+\log e)}{\textstyle\bigl(\vec x,\ET_ix_i,\dots,\ET_ix_i\bigr)},\] where
$r$ is the sequence of random coin tosses that determines the leaves of the formula. Notice that
$\land\in\cln{MPT}^k_0=[\theta^{k+1}_k]$, hence $\ET_ix_i$ can be easily expressed by a logspace-computable sequence of
$O(\log n)$-depth $\{\theta^{k+1}_k\}$-formulas. It is also straightforward to compute $F_{N,d}$ in logspace
given $N$, $d$, and the random sequence~$r$.

Finally, to compute $N=\fl{\sigma^{-1}t}$, the algebraic number~$\sigma^{-1}$ can be approximated by a $\tc$~function
(hence a logspace function) by~\cite{ej:polyroot}, but since we actually need it only with $O(\log n)$ bits of
precision, even a brute force approach is sufficient: we can evaluate the polynomial $x^k-(k+1)t^{k-1}x+kt^k$ (which is
$t^kh(xt^{-1})$) in parallel for each $x=n,n+1,\dots,2t$, and let $N$ be the argument where it switches sign from
negative to positive. Thus, $T_{n,t,e,r}$ is computable by a logspace function.
\end{Pf}
\begin{Rem}\label{rem:alpha-k}
The expansion~\eqref{eq:8} easily implies that for large~$k$, the constant $\sigma_k$ from Theorem~\ref{thm:rand-th-fla} is
$1-2k^{-2}+O(k^{-3})$.
\end{Rem}

\begin{Thm}\label{thm:b-cmp-rand-tptc}
If $[B]\Sset\cln{PT}^k_\alpha$ for some $k\in\N$ and $\alpha\in\two$, then $B$-$\cmp$ is $\tpt$-complete under
randomized logspace reductions. More precisely, for any language $L\in\tpt$, there exists a constant~$c$ and a
logspace function $R(w,e,r)$ with $e$ given in unary such that for all strings $w$ of length~$n$, and for all~$e$,
\begin{align}
\label{eq:11}w\in L&\implies\PR_{\lh r=(n+e)^c}[R(w,e,r)\in B\text-\cmp]=1,\\
\label{eq:14}w\notin L&\implies\PR_{\lh r=(n+e)^c}[R(w,e,r)\in B\text-\cmp]\le2^{-e}.
\end{align}
\end{Thm}
\begin{Pf}
We will assume $[B]\Sset\cln T^k_0$: we may pass from $\cln T^k_0$ to $\cln{PT}^k_0$ in the same way as in the proof of
Theorem~\ref{thm:p-cmp-th2}, and the case of $[B]\Sset\cln{PT}^k_1$ is dual. Without loss of generality, $k\ge3$.

We use the reduction from Lemma~\ref{lem:th-seq} and Theorem~\ref{thm:cmp-main} with the following two modifications:
\begin{enumerate}
\item\label{item:11}
We express the formulas $x_i\land\fii_i$ by $B$-formulas.
\item\label{item:12}
In place of the threshold function~$\theta^m_t$, we use the randomly generated formula $T_{m,t,e,r}$ from
Theorem~\ref{thm:rand-th-fla}.
\end{enumerate}
As for~\ref{item:11}, recall that the formulas $\fii_i$ supplied by Lemma~\ref{lem:thp2-compl} are CNFs, hence they may be
arranged to have depth $O(\log n)$. We may assume them to be written in the $\{\land,\neg\}$ basis. We then write
$x_i\land\fii_i$ in the basis $\{\land,\nto\}\sset\cln T^\infty_0$ by replacing each subformula $\neg\psi$ with
$x_i\land\neg\psi$. Since $\cln T^\infty_0\sset[B]$, we may rewrite the formulas as $B$-formulas by Lemma~\ref{lem:conv}.

Concerning \ref{item:12}, notice first that in Lemma~\ref{lem:th-seq}, we may assume $n$ to be sufficiently large, and the
parameters we take are $m\approx n^2$, $m-t\approx n$, thus $t>\sigma_km$, justifying the use of
Theorem~\ref{thm:rand-th-fla}. We again rewrite the formulas as $B$-formulas using Lemma~\ref{lem:conv}. Crucially, we have to
use the same $T_{m,t,e,r}$~formula for constructing both $f_{\mathrm{even}}$ and $f_{\mathrm{odd}}$.

It is clear from the construction that \eqref{eq:14} holds, but we need more work to establish~\eqref{eq:11}, as it is
not obvious that it holds with no error. If $w\in L$, let $\p{\fii_i:i<2n}$ and $j$ be as in Theorem~\ref{thm:cmp-main}, so
that $j$ is even, and put $s=j/2$. Then in the definition of both $f_{\mathrm{even}}=f_{\fii_0,\fii_2,\dots,\fii_{2n-2}}$
and $f_{\mathrm{odd}}=f_{\fii_1,\fii_3,\dots\fii_{2n-1}}$, the first $s$ of the $\fii_i$ formulas are satisfiable, and the
rest are unsatisfiable. Going back to Lemma~\ref{lem:th-seq}, let us abbreviate by $T(x_0,\dots,x_{m-1})$ the formula
$T_{m,t,e,r}$ we use in place of~$\theta^m_t$. Then
\[f_{\mathrm{even}}\equiv T(x_0\land\fii_0,x_1\land\fii_2,\dots,x_{s-1}\land\fii_{2s-2},\underbrace{0,\dots,0}_{n-s},x_n,\dots,x_{m-1})\]
and
\[[\nto,f_{\mathrm{even}}]=[\nto,T(x_0,\dots,x_{s-1},0,\dots,0,x_n,\dots,x_{m-1})]\]
by the argument in Lemma~\ref{lem:th-seq} (avoiding renumbering of variables or permuting the arguments of~$T$). Since the
same~$T$ was also used to construct $f_{\mathrm{odd}}$, we obtain likewise
\[[\nto,f_{\mathrm{odd}}]=[\nto,T(x_0,\dots,x_{s-1},0,\dots,0,x_n,\dots,x_{m-1})],\]
hence $[\nto,f_{\mathrm{even}}]=[\nto,f_{\mathrm{odd}}]$.
\end{Pf}

Notice that if the language $L$ we are reducing to $B$-$\cmp$ is in~$\cxt{BH}$, the number $n$ in the proof of
Theorem~\ref{thm:cmp-main} can be taken as constant, hence also $m$ and~$t$ in Lemma~\ref{lem:th-seq} are constant, and we may
just fix a representation of $\theta^m_t$ by a $B$-formula in advance, avoiding the complicated randomized construction
from Theorem~\ref{thm:rand-th-fla}.
\begin{Cor}\label{cor:b-cmp-bh-hard}
If $[B]\Sset\cln{PT}^k_\alpha$ for some $k\in\N$ and $\alpha\in\two$, then $B$-$\cmp$ is $\cxt{BH}$-hard.
\noproof\end{Cor}

\begin{figure}[t]
\centering
\includegraphics[width=300bp]{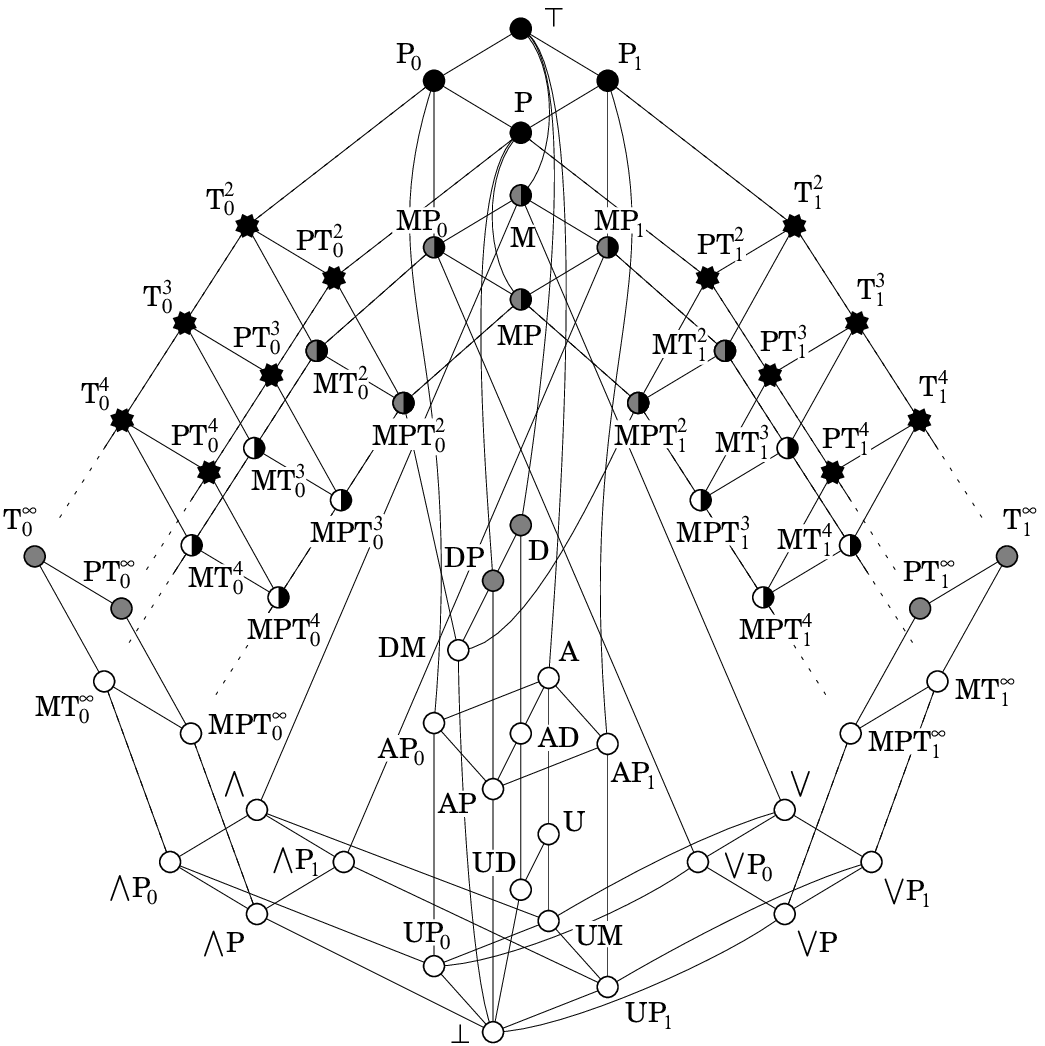}
\caption{The complexity of $B\text-\cmp$
(\legend{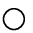}:~in~$\ptime$;
\legend{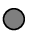}:~$\cxt{coDP}$-complete;
\legend{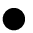}:~$\tpt$-complete;\hfil\break
\legend{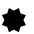}:~$\tpt$-complete under randomized reductions;
\legend{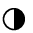}:~in~$\tpt$;
\legend{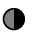}:~in~$\tpt$, $\cxt{coDP}$-hard).%
}
\label{fig:cmp}
\end{figure}
Let us summarize the results of Section~\ref{sec:restr-input-bases} (see Fig.~\ref{fig:cmp}):
\begin{Cor}\label{cor:b-cmp-summ}
Let $B\sset\Op$ be finite.
\begin{enumerate}
\item If $B\sset\cln{MT}^\infty_0$, $\cln{MT}^\infty_1$, $\cln\ET$, $\cln\LOR$, $\cln A$,
or~$\cln{DM}$, then $B\text-\cmp\in\ptime$.
\item If  $[B]$ is $\cln T^\infty_0$, $\cln{PT}^\infty_0$, $\cln T^\infty_1$, $\cln{PT}^\infty_1$, $\cln D$, or
$\cln{DP}$, then $B$-$\cmp$ is $\cxt{coDP}$-complete.
\item If $[B]\Sset\cln{PT}^k_\alpha$ for some $k\in\N$ and $\alpha\in\two$, then $B$-$\cmp$ is $\tpt$-complete under
randomized reductions, and even under deterministic reductions if $[B]\Sset\cln P$. Also, $B$-$\cmp$ is
$\cxt{BH}$-hard.
\item If $\cln{MPT}^k_\alpha\sset[B]\sset\cln M$ for some $k\in\N$ and $\alpha\in\two$, then $B$-$\cmp$ is in $\tpt$.
If $[B]\Sset\cln{MPT}^2_\alpha$, $B$-$\cmp$ is $\cxt{coDP}$-hard.
\noproof
\end{enumerate}
\end{Cor}

While randomized reductions are a nuisance, the real problem is the last item of Corollary~\ref{cor:b-cmp-summ}, where
the upper and lower bounds (if any) are far from each other. Notice that since any non-constant monotone function is
both $0$- and $1$-preserving, $\cln M$-$\cmp$ is logspace-equivalent to $\cln{MP}$-$\cmp$, and (using also duality)
for any $k\ge1$, the problems $\cln{MT}^k_0$-$\cmp$, $\cln{MPT}^k_0$-$\cmp$,  $\cln{MT}^k_1$-$\cmp$, and
$\cln{MPT}^k_1$-$\cmp$ are logspace-equivalent.
\begin{Prob}\label{prob:b-cmp-mt}
What is the complexity of $B$-$\cmp$ for $\cln{MPT}^k_\alpha\sset[B]\sset\cln M$? 
\end{Prob}

Our first hunch is that all these problems should be $\tpt$-complete just like their non-monotone versions, but on
second thought, it is conceivable that, for example, we can learn some properties of monotone functions by a randomized
process such as in the proof of Theorem~\ref{thm:rand-th-fla}, hence the expected answer is not as clear-cut.

We note that B\"ohler and Schnoor~\cite{boeh-schn} left open a similar problem about the complexity of certain cases of $B$-$\cmp_C$.

\section{Conclusion}\label{sec:conclusion}

We have undertaken a thorough investigation of the complexity of the Boolean clone membership problem $\cmp$ and its
variants. Most importantly, we proved that $\cmp$ is $\tpt$-complete, and in particular, strictly harder than
any of the fixed-clone problems $\cmp_C$ or $\cmp^f$, barring collapse of the polynomial hierarchy.

Moreover, we obtained a representative (even if incomplete) picture of how the complexity depends on the basis~$B$ of
gates allowed in the input. As expected, it shows a major dividing line depending on whether $[B]$ has finitely many
subclones: in the latter case the complexity drops down inside the Boolean hierarchy---in fact, to $\cxt{coDP}$.
However, there seems to be also a more subtle dividing line based on whether $B$ consists of monotone functions only:
in the finite subclone case, this makes the complexity of $B$-$\cmp$ go further down to~$\ptime$ (though this also
happens for some non-monotone cases, namely when $B$ consists of affine functions); in the infinite subclone case, it
separates the area of more-or-less $\tpt$-complete instances of $B$-$\cmp$ from a terra incognita.

\subsection*{Acknowledgements}

I would like to thank the anonymous referees for helpful comments.

The research was supported by grant 19-05497S of GA \v CR. The Institute of Mathematics of the Czech Academy of
Sciences is supported by RVO: 67985840.

\bibliographystyle{mybib}
\bibliography{clonememb}
\end{document}